\documentclass[showpacs,aps,prd,nofootinbib,showkeys,superscriptaddress,twocolumn]{revtex4}
\usepackage{graphicx}
\usepackage{bm}
\usepackage{bbold}
\usepackage{amssymb}
\usepackage{amsmath,latexsym}
\usepackage{amsfonts}
\usepackage[usenames]{color}
\usepackage{subfigure}
\usepackage{subfigure}
\usepackage{slashed}
\usepackage{multirow,array}
\usepackage{mathtools}
\usepackage{mathrsfs}
\usepackage{dsfont}
\usepackage[colorlinks=false,linktocpage=true]{hyperref}
\usepackage{hyperref}
\usepackage[utf8]{inputenc}
\usepackage{aurical}
\usepackage[T1]{fontenc}
\usepackage{multirow}
\newcommand{\be}{\begin{equation}}
\newcommand{\ee}{\end{equation}}
\newcommand{\bea}{\begin{eqnarray}}
\newcommand{\eea}{\end{eqnarray}}
\newcommand{\beal}{\begin{align}}
\newcommand{\enal}{\end{align}}
\newcommand{\bs}{\begin{subequations}}
\newcommand{\es}{\end{subequations}}
\newcommand{\besp}{\begin{split}}
\newcommand{\eesp}{\end{split}}

\newcommand{\g}{g_{\mu\nu}}
\newcommand{\gup}{g^{\mu\nu}}
\newcommand{\h}{h_{\mu\nu}}
\newcommand{\hij}{h^{ij}}
\newcommand{\dij}{\delta^{ij}}

\newcommand{\e}{\eta_{\mu\nu}}

\newcommand{\Du}{\Delta^{\mu\nu}}

\newcommand{\pro}{\bar{p}}
\newcommand{\bd}{\bar{n}}
\newcommand{\um}{\bar{u}^\mu}

\newcommand{\du}{\delta u}
\newcommand{\tmn}{T^{\mu\nu}}

\newcommand{\jm}{J^\mu}

\newcommand{\fmn}{F^{\mu\nu}}

\newcommand{\ene}{\bar{\epsilon}}
\newcommand{\den}{\delta\epsilon}
\newcommand{\dep}{\delta p}
\newcommand{\dbd}{\delta n}

\newcommand{\am}{A^\mu}

\newcommand{\smn}{\sigma^{\mu\nu}}
\newcommand{\Smn}{S^{\mu\nu}}
\newcommand{\I}{I^\mu}

\newcommand{\mO}{\mathcal{O}}
\newcommand{\bk}{\bold{k}}
\newcommand{\hbk}{\hat{\bk}}
\newcommand{\bx}{\bold{x}}

\newcommand{\gL}{\frg_L}

\newcommand{\mc}[1]{\mathcal{#1}}

\newcommand{\frg}{\text{$\mathfrak{g}$}}
\newcommand{\eva}{\lambda_{i}}
\newcommand{\etr}[1]{\hat{e}_{\bold{T}_{#1}}}
\newcommand{\dq}{\delta q}
\newcommand{\Hb}{\bar{H}}
\newcommand{\geta}{\gamma_{\eta_0}}

\newcommand{\C}{\mc C}

\newcommand{\bw}{\bar{w}}
\newcommand{\ubk}{\bk^{\perp}}
\newcommand{\ufrg}{\frg^{\perp}}
\newcommand{\bbk}{\bm{K}}
\newcommand{\hbbk}{\hat{\bbk}}

\newcommand{\overbar}[1]{\mkern 1.5mu\overline{\mkern-1.5mu#1\mkern-1.5mu}\mkern 1.5mu}

\begin{document}
\interfootnotelinepenalty=10000
\title{Stochastic hydrodynamics and long time tails of an expanding conformal charged fluid}
\date{\today}

\author{M.~Martinez}
\affiliation{Department of Physics, North Carolina State University, Raleigh, NC 27695}

\author{T.~Sch{\"a}fer}
\affiliation{Department of Physics, North Carolina State University, Raleigh, NC 27695}

\begin{abstract}
We investigate the impact of hydrodynamic fluctuations on correlation functions in a scale invariant fluid with a conserved $U(1)$ charge. The kinetic equations for the two-point functions of pressure, momentum and heat energy densities are derived within the framework of stochastic hydrodynamics. The leading non-analytic contributions to the energy-momentum tensor as well as the $U(1)$ current are determined from the solutions to these kinetic equations. In the case of a static homogeneous background we show that the long time tails obtained from hydro-kinetic equations reproduce the one-loop results derived from statistical field theory. We use these results to establish bounds on transport coefficients. We generalize the stochastic equation to a background flow undergoing Bjorken expansion. We compute the leading fractional power $\mc O((\tau T)^{-3/2})$ correction to the $U(1)$ current and compare with the first order gradient term. 
\end{abstract}


\maketitle

\section{Introduction}
\label{sec:intr}

The hydrodynamic description of relativistic heavy ion collisions at the Relativistic Heavy Ion Collider (RHIC) and the Large Hadron Collider (LHC) has been extremely successful \cite{Romatschke:2017ejr,Jeon:2015dfa,Teaney:2009qa}. State-of-the-art models contain not only higher order gradient terms in relativistic fluid dynamics, but also final state kinetic theory afterburners, and initial state models that account for event-by-event fluctuations. There is strong evidence for the role of initial state fluctuations from the observation of odd azimuthal moments of hydrodynamic flow \cite{Alver:2010gr}.

The role of thermodynamic fluctuations during the evolution of the system is much less explored. These fluctuations arise from the fact that fluid dynamics is a coarse-grained description, so that at any coarse graining scale unresolved microscopic degrees of freedom lead to fluctuations of the macroscopic variables. These effects are interesting, because hydrodynamics is a non-linear theory, so that couplings between hydrodynamic modes lead to novel phenomena beyond just Gaussian fluctuations in the hydrodynamic variables. For example, it is known that fluctuations lead to hydrodynamic "tails"~\cite{pomeau1975time,fox1978gaussian}, which are non-analytic contributions to the time (or frequency) dependence of correlation functions \cite{Kovtun:2012rj,Chafin:2012eq,Martinez:2017jjf}. In the gradient expansion these terms are formally more important than corrections beyond the Navier-Stokes approximation. The relative importance of fluctuations depends on the value of the transport coefficients. Fluctuations are enhanced in low viscosity fluids, such as the quark gluon plasma, and they are suppressed in the large N limit, where in the case of gauge theories N is the number of colors~\cite{Chafin:2012eq,Kovtun:2005ev}.

More importantly, hydrodynamic fluctuations dominate the dynamics near a critical point, and any fully dynamical description of fluctuations of conserved charges in a heavy ion collision must include fluctuations  in the hydrodynamic evolution equations. This implies that the study of hydrodynamic fluctuations is crucial for interpreting the results from the RHIC beam energy scan program.

There are a variety of techniques for taking int account hydrodynamic fluctuations. In homogeneous systems fluctuations have been studied using diagrammatic or effective action methods \cite{Martin:1973zz,Kovtun:2012rj}. Combined with the epsilon expansion, these methods have led to the determination of dynamical critical exponents \cite{Hohenberg:1977ym}. In a complicated environment like a heavy ion collision these methods are less convenient. An alternative is to numerically study stochastic hydrodynamic equations. There is some progress in this effort \cite{Nahrgang:2018afz}, but the calculations are difficult, and the results are not always simple to interpret. Several authors have derived deterministic equations for the time evolution of hydrodynamic $n$-point functions \cite{Akamatsu:2016llw,Stephanov:2017ghc,Mukherjee:2015swa}. These methods rely on linearizing the fluid dynamic equations, and on truncations in the number of moments, but the resulting equations are easier to evolve in time, and the comparison to analytical results for homogeneous systems is more straightforward.

In this work we adopt the latter approach, extending a formalism developed by Akamatsu et al.~\cite{Akamatsu:2016llw} to two-point functions involving a conserved number density. This is an important problem, because the conserved baryon density (or the entropy per baryon number) is believed to be a order parameter for a possible critical point in the QCD phase diagram~\cite{Son:2004iv}. In the present work, we will restrict ourselves to the two-point function of a scale invariant fluid, leaving the role of a critical equation of state to a future study. Estimates of the role of critical fluctuations have recently appeared in \cite{Akamatsu:2018vjr}. As indicated above we will linearize the equations of motion, an expand in powers of fluctuating fields and external sources. We provide bounds on the diffusion constant, and derive the tails of diffusive correlators in a Bjorken backgound. 

We use the following conventions: We use a bar to denote background values of the hydrodynamic variables. Four dimensional indices are denoted by Greek letters $\mu,\nu,...=0,1,2,3$; three dimensional indices are labeled by Latin letters $i,j,...=1,2,3$. Three dimensional spatial vectors are denoted by bold letters ($\bold{x}$,$\bold{k}$). The signature of the metric is taken to be `mostly-plus'. A four dimensional vector is denoted by $x^\mu=(x^0,\bold{x})$. The fluid velocity is a time-like vector with $u_\mu u^\mu=-1$. We denote the spatial measure of the integral $\int d^3\bx\sqrt{-g}\equiv\int_{\bx}$ while the three dimensional measure in momentum space is $\int d^3\bk/[(2\pi)^3\sqrt{-g}]\equiv\int_{\bk}$.

\section{Relativistic Hydrodynamic fluctuations of a conformal charged fluid}
\label{sec:constrel}

In this section we describe the theoretical framework for studying the evolution of hydrodynamic fluctuations on top of an evolving background for a charged fluid. We consider a relativistic fluid with an unbroken global $U(1)$ symmetry (such as baryon number) where parity is conserved. Our goal is to determine the response functions via the variational approach and thus, as we introduce external fictitious sources which excite a particular set of hydrodynamic modes. For a more detailed explanation of this method we refer to the reader to Ref.~\cite{Kovtun:2012rj}. 

The local conservation laws for this system are~\cite{Weinberg:1972kfs,landaufluid}
\bs
\label{eq:conslaw}
\beal
D_\mu\tmn&= F^{\nu\lambda} J_\lambda\,,\\
D_\mu\jm &=0\,,
\end{align}
\es
where $\tmn$ is the energy-momentum tensor of the fluid, $\jm$ is the $U(1)$ current and $\fmn=\partial^{[\mu}A^{\nu]}$ is the field strength tensor of a gauge field $\am$ coupled to the conserved current. The energy momentum tensor couples to the metric $\gup$, and the covariant derivative associated with $\gup$ is denoted by $D_\mu$. 

In the Landau frame where $u_\mu\tmn=\epsilon u^\nu$ (being $u^\mu$ the time-like vector denoting the fluid velocity and $\epsilon$ the energy density) the energy-momentum tensor $\tmn$ and the current $\jm$ can be decomposed as
\begin{subequations}
\label{eq:constrel}
\begin{align}
\label{eq:tmndecom}
\tmn&=\epsilon\,u^\mu u^\nu\,+\,p\,\Delta^{\mu\nu}\,+\,\pi^{\mu\nu}\,+\,\Smn\,,\\
\label{eq:jmdecom}
\jm&=\,n u^\mu+\,v^\mu\,+\,\I\,.
\end{align}
\end{subequations}
In the previous equations the macroscopic quantities of the fluid are the isotropic pressure $p$, the viscous stress tensor $\pi^{\mu\nu}$, the particle density $n$ and the diffusive current $v^\mu$. Fluctuating contributions to the energy-momentum tensor and current are denoted by $\Smn$ and $\I$. We introduce the tensor $\Delta^{\mu\nu}=\gup\,+\,u^\mu u^\nu$ which is orthogonal to the fluid velocity. 

In order to solve the conservation laws of stochastic hydrodynamics~\eqref{eq:conslaw} one requires an equation of state and the constitutive relations that express the conserved currents in terms of hydrodynamic variables. Within the Navier-Stokes approximation and for conformal systems the constitutive relations for the stress tensor $\pi^{\mu\nu}$ and the particle diffusion current $v^\mu$ read as respectively
\begin{subequations}
\label{eq:tvquant}
\begin{align}
\label{eq:pi}
\pi^{\mu\nu}&=-\eta_0\smn\,,\\
\label{eq:vmu}
v^\mu&=\sigma_0\,\Delta^{\mu\nu}\left[E_\nu\,-\,T\,\partial_\nu\left(\frac{\mu}{T}\right)\right]\,,
\end{align}
\end{subequations}
where $\smn$ is the Navier-Stokes shear tensor and $E^\mu$ is an external electric field. $\smn$ and $E^\mu$ are defined as follows
\begin{subequations}
\begin{align}
\label{eq:electric}
E^\mu&=\fmn u_\nu\,,\\
\label{eq:streten}
\smn&=2\,D^{\langle\mu}u^{\nu\rangle}\,,
\end{align}
\end{subequations}
where $W^{\langle\mu\nu\rangle}=\Delta^{\mu\nu}_{\alpha\beta}A^{\alpha\beta}$ with 
\be
\Delta^{\mu\nu}_{\alpha\beta}=
 \left[\Delta^\mu_\alpha\Delta^\nu_\beta
  +\Delta^\nu_\alpha\Delta^\mu_\beta
  -(2/3)\Delta^{\mu\nu}\Delta_{\alpha\beta}\right]/2.
\ee
Note that $\Delta^{\mu\nu}_{\alpha\beta}$ is a 2-rank tensor which is traceless, symmetric, and orthogonal to the fluid velocity $u^\mu$. The bare transport coefficients associated with charge conductivity and shear viscosity are denoted by $\sigma_0$ and $\eta_0$ respectively. In Eqs.~\eqref{eq:tvquant} we introduce the projector orthogonal to the fluid velocity $\Du=\gup+u^\mu u^\nu$. 

The set of equations~\eqref{eq:conslaw} are not complete unless we specify the equation of state as well as the functional form of the random sources $\Smn$ and $\I$. At leading order in the low momentum (gradient) expansion we can model the random sources as delta-correlated white noise~\cite{landauv9,landaufluct,Kapusta:2011gt}, i.e., 
\begin{subequations}
\label{eq:noisevar}
\begin{align}
\label{eq:tmnvar}
\langle\,S^{\mu\nu}(x^0_1,\bx_1)\,S^{\alpha\beta}(x^0_2,\bx_2)\,\rangle&=\,2T\,\left[2\,\eta_0\,\Delta^{\mu\nu,\alpha\beta}\right]\,\nonumber\\&\hspace{-1.5cm}\times \delta\left(x^0_1-x^0_2\right)\,\delta^{(3)}\left(\bx_1-\bx_2\right)\,,\\
\label{eq:jmvar}
\langle I^{\mu}(x^0_1,\bx_1)\,\,I^\nu(x^0_2,\bx_2)\,\rangle&=\,2T\,\sigma_0\,\Delta^{\mu\nu}\,\nonumber\\
&\hspace{-1.5cm}\delta\left(x^0_1-x^0_2\right)\,\delta^{(3)}\left(\bx_1-\bx_2\right)\,,\\
\label{eq:null}
\langle\,S^{\mu\nu}(x^0_1,\bx_1)\,I^\lambda(x^0_2,\bx_2)\,\rangle&=0\,.
\end{align}
\end{subequations}
In the leading order of the perturbation each hydrodynamical field entering in the RHS of the previous equations is replaced by its mean field value. These approximations will allow us to recast the problem of solving the equations of stochastic hydrodynamics onto a set of coupled Langevin equations. 

In this work we consider a conformally invariant fluid whose equation of state (EOS) is
\be
\label{eq:confeos}
p(T,\mu)=T^4\,g(\mu/T)\,, 
\ee
where $g$ is an arbitrary function. For this particular equation of state, the conformal symmetry implies the ratio $\mu/T$ is invariant under Weyl rescaling~\cite{Erdmenger:2008rm}. Different thermodynamic relations associated to the EOS~\eqref{eq:confeos} are listed in App.~\ref{app:thermo}.

\section{Hydrodynamic fluctuations around a static background}
\label{sec:hydrofluc}

Hydrodynamic tails appear in the long time behavior of response functions. We compute these response functions by coupling the hydrodynamic evolution in the presence of noise terms to external fields, and then determine the variational derivative of currents with respect to the sources. Specifically, the metric can be used to study response functions of the stress tensor, and the external gauge field determines the current response. 
 
In order to study diffusive terms in the shear response in a static background it is sufficient to consider a metric of the type $\g=\e+\h$ where the metric perturbation is parametrized as~\cite{Akamatsu:2016llw}
\be
\label{eq:metans}
h_{ij}=\text{diag}\,(1,1,-2)\,h_s(x^0)\,,
\ee
where $h_s$ is a time-dependent scalar function. The current response is studied by considering an external gauge field of the form 
\be
A^\mu=(0,\delta A^i(x^0)).
\ee
In a static background the hydrodynamic equations for the background hydrodynamic are
\be
\label{eq:backeos}
\partial_0\ene=0\,,\hspace{1cm}\partial_0\bd=0\,.
\ee
The solutions of these two equations are given in terms of static background configurations, i.e., $\ene=\epsilon_0$ and $\bd=n_0$ respectively. In the following we will also make use of the heat energy density \cite{kadanoffmartin,forster1995hydrodynamic}
\be
q=\ene-\bar{w}\,\bd\, , \hspace{0.4cm}
\bar{w}=(\ene+\bar{p})/\bd\, , 
\ee
where $\bar{w}$ is the enthalpy per particle. From the first law of thermodynamics the variation of the heat energy density represents the change of the entropy per particle~\footnote{Eq.~\eqref{eq:heaten} considers that the total number of particles $N$ is fix and thus, all the thermodynamic derivatives must be taken at fixed $N$~\cite{kadanoffmartin,forster1995hydrodynamic}.}
\be
\label{eq:heaten}
\dq\equiv T\,\bd \,\delta\left(\frac{s}{n}\right)\,.
\ee
Given the conservation of the entropy (or the heat energy density) one can define the relation between the pressure and the energy density as 
\be
\label{eq:adspeed}
v_a^2=\frac{\partial p}{\partial\epsilon}\biggl|_{s/n}\,
=\frac{\partial p}{\partial\epsilon}\biggl|_{n}\,
 +\,\frac{1}{w}\,\frac{\partial p}{\partial n}\biggl|_\epsilon
 \equiv\,\beta_1\,+\,\frac{1}{w}\,\beta_2\,,
\ee
where we recognize $v_a^2$ as the (adiabatic) speed of sound. For the EOS~\eqref{eq:confeos}  $\beta_1=1/3$, $\beta_2=0$ so $v_a^2=1/3$  as expected (see App.~\ref{app:thermo}).

\subsection{The Navier Stokes-Langevin equations}
\label{subsec:L-NS}

In this section we will derive the evolution equation for hydrodynamic fluctuations in the presence of noise, the linearized Navier-Stokes Langevin equation. We begin by linearizing the equations of motion of relativistic fluid dynamics around a static background. We write the hydrodynamic variables as 
\be
\label{eq:pert}
\begin{split}
&\epsilon=\ene+\den\,,\hspace{0.5cm} p=\pro + \dep\,,\hspace{0.5cm} n=\bd+\dbd\,,\\
&u^\mu=\um+\du^\mu\,,\hspace{1cm}\um=(1,\bold{0})\,.
\end{split}
\ee
where $\den$, $\dep$ and $\dbd$ are perturbations in the energy density, pressure and particle density, respectively. We will write the equations of motion in a mixed representation where we perform a Fourier transform with respect to spatial variables\footnote{The spatial Fourier transform is defined as
\be
 B\equiv B(x^0,\bk)=\int_{\bx}\,e^{-i\bold{k}\cdot\bold{x}}B(x^0,\bold{x})\,.
\ee
}.
The linearized stochastic hydrodynamic equations are
\begin{subequations}
\label{eq:linstheos}
\begin{align}
\label{eq:linstheos-ene}
&\partial_0\den+\,i\,\bar{H}\,k_i\,\delta u_i\,=\,-\bd\,E_i\,\delta u_i\,,\\
\label{eq:linstheos-gi}
&\bar{H}\,\partial_0 \delta u_j \,+\,i\,k_j\dep\,+\,\geta\,\bar{H}\,\left[k^2\,\delta_{ji}\delta u_i\,+\,\frac{1}{3}\,k_j\,k_i\,\delta u_i\right]\nonumber\\
&\hspace{1.5cm}
=E_j\,\dbd\,-\,\frac{\bar{H}\,}{2}\delta u_k\,\partial_0h_{jk}\,-i\,k_i\,S_{i j}\,,\\
\label{eq:linstheos-part}
&\partial_0\dbd\,+\,i\,\bd\,k_i\,\delta u_i\,+\,\sigma_0\,k^2\left[\alpha_1\,\den\,+\,\alpha_2\,\dbd\right]\nonumber\\
&\hspace{1.5cm}
=-\,i\,k_i\,I_i\, . 
\end{align}
\end{subequations}
Here, we have defined the electric field by $E^{i}(x^0,\bk) = F^{i0}\bar{u}_0 = F^{0i}$, and we denote $k^2=|\bold{k}|^2$. We have also introduced the momentum diffusion constant
\be
\geta=\frac{\eta_0}{\bar{H}}\, , \hspace{0.5cm}
\bar{H}=\ene+\bar{p}\, , 
\ee
where $\bar{H}$ is the enthalpy density. In Eq.~\eqref{eq:linstheos-part} we did not keep higher order terms in frequency and wave number. We have also defined the thermodynamic quantities
\bs
\label{eq:thermcoeff}
\begin{align}
\label{eq:alpha1}
\alpha_1&=\left(\frac{\partial\,\mu}{\partial\,\epsilon}\right)_n-\frac{\mu }{T}\,\left(\frac{\partial\,T}{\partial\,\epsilon}\right)_n\,,\\
\label{eq:alpha2}
\alpha_2&=\left(\frac{\partial\,\mu}{\partial\,n}\right)_\epsilon-\frac{\mu }{T}\,\left(\frac{\partial\,T}{\partial\,n}\right)_\epsilon\,.
\end{align}
\es
The natural degrees of freedom of Eqs.~\eqref{eq:linstheos} are the variations of the energy density $\den$, particle density $\dbd$ and fluid velocity~$\delta u^i$. However we are free to use any other set of variables when analyzing Eqs.~\eqref{eq:linstheos}. It is more convenient to use the heat energy density $\dq$, pressure $\dep/v_a$ and momentum flux density $\frg_i\equiv T_{0i}$. These variables have the property that in thermal equilibrium their fluctuations are statistically independent~\cite{landauv9,kadanoffmartin,forster1995hydrodynamic}. For the conformal EOS~\eqref{eq:confeos} and close to the thermal equilibrium we have $\varphi_a=M_{ab}\psi_b$ with $\varphi=(\den,\delta u_i,\dbd)$ and $\psi=(\dep/v_a,\frg_i,\dq)$ and the matrix $M_{ab}$ given by
\be
\label{eq:thermo}
M=
\begin{pmatrix}
\frac{1}{v_a}&0&0\\
0&(\Hb)^{-1}\delta_{ji}&0\\
\frac{1}{\bar{w}v_a}&0&-\frac{1}{\bar{w}}
\end{pmatrix}\,.
\ee
The momentum density is a three-dimensional vector. It is convenient to express this vector in a basis spanned by the vectors $\hbk=\bk/k$ and $\hat{e}_{\bold{T}_{i}}$ (i=1,2) defined by~\cite{Akamatsu:2016llw}
\bs
\label{eq:3d-spbasis}
\beal
\hbk&=\left(\sin\theta\cos\phi,\sin\theta\sin\phi,\cos\theta\right)\,,\\
\etr 1&=\left(-\sin\phi,\cos\phi,0\right)\,,\\
\etr 2&=\left(\cos\theta\cos\phi,\cos\theta\sin\phi,-\sin\theta\right)\,.
\end{align}
\es
In this basis the momentum flux density $\frg_i$ is decomposed as
\begin{equation}
\label{eq:gsplit}
\frg=\hbk\,\underbrace{\left(\hbk\cdot \frg\right)}_{\equiv \gL}\,+\,\sum_{l=1}^2\left(\etr l\right)\,\underbrace{\left(\etr l\cdot \frg\right)}_{\frg_{\bm{T}_l}}\,.
\end{equation}
Using~\eqref{eq:thermo} and~\eqref{eq:linstheos} we obtain the Navier-Stokes-Langevin (NSL) equations in the form
\be
\label{eq:NSLeqs}
\partial_0\psi^a_\bk\,+\,\left(\mc A^{ab}\,+\,\mc D^{ab}\,\right)\psi^b_\bk=\,\mc P^{ab}\,\psi^b_\bk\,+\,\,\xi^a\,,
\ee
where 
\be
\psi^a_\bk=(\dep/v_a,\frg_i,\dq) \;\;\; (i=1,2,3) 
\ee
is a five dimensional vector representing the fluctuating hydrodynamic fields. The acoustic and diffusive matrices, denoted by $\mc A$ and $\mc D$ respectively, are
\bs
\label{eq:Rmatstat}
\begin{align}
\label{eq:accoustic}
\mc A
&=\,k\,
\begin{pmatrix}
0 &i\,v_a\,\hbk_i&0\\
i\,v_a\,\hbk_j&0&0\\
0&0&0
\end{pmatrix}\,,\\
\label{eq:diffusive}
\mc D&=k^2\,\text{diag.}\left(0,\geta\left(\delta_{ij}+\frac{1}{3}\hbk_i\hbk_j\right),D_0\right)\,,
\end{align}
\es
where we have defined the diffusion coefficient $D_0=\sigma_0\,\alpha_2$. The NSL equation describes five hydrodynamic modes, and the matrices $\mc A$ and $\mc D$ determine the dispersion relation of these modes at $\mO(k)$ and $\mO(k^2)$ in a expansion the wave number $k$.

At leading order (LO) the acoustic matrix $\mc A$ describes two sound modes with eigenvalues $\lambda_{\pm}=\pm\,i\,v_a$, one diffusive (d) mode and two shear modes ($\bm{T}_1,\bm{T}_2$)~\cite{kadanoffmartin,forster1995hydrodynamic,chaikin2000principles,Kovtun:2012rj}. At this order in the expansion the eigenvalues of the diffusive and shear modes are $\lambda_A=0$ ($A=d,\bm{T}_1,\bm{T}_2$). The absence of dissipation at LO implies that $\dq$ is a conserved quantity~\cite{forster1995hydrodynamic}. 

At next-to-leading order (NLO) non-diagonal terms in the matrix $\mc D$ lead to mixing between the pressure, the heat energy density, and the longitudinal momentum density. Both the momentum density and the heat energy receive real (diffusive) corrections $\mO(k^2)$. As a result the dynamics of the modes $(\bm{T}_1,\bm{T}_2)$ remains purely diffusive, while $(\dep/v_a,\gL,\dq)$ exhibit a combination of diffusive and reactive behaviour. 

The source matrix $\mathcal{P}^{ab}$ in Eq.~\eqref{eq:NSLeqs} takes into account the non-linear couplings between the fluctuating hydrodynamic fields and the external sources. Its explicit components are 
\be
\label{eq:Rmatrix}
\mathcal{P}=
\begin{pmatrix}
0&-\frac{v_a}{\bar{w}}\,E_j&0\\
\frac{1}{\bar{w}\,v_a}\,E_j&-\frac{\partial_0 h_{kj}}{2}&-\frac{1}{\bar{w}}\,E_j\\
0&-\frac{1}{\bar{w}}E_j&0
\end{pmatrix}
\ee
Finally the vector $\xi^a$ carries the information of the random noise sources. Its components are
\bs
\label{eq:noiseonepoint}
\beal
\xi^a=\begin{pmatrix}
0\\
-i\,k_i\,S_{ij}\\
i\,\bar{w}\,k_i\,I_i
\end{pmatrix}
\end{align}
\es
The solutions of the NSL equations~\eqref{eq:NSLeqs} determine the linear evolution of variations of the pressure, heat energy, and momentum flux densities in response to noise and external sources.

\section{Evolution equations for two-point correlation functions}
\label{sec:sol}

The general solution of the NSL equations~\eqref{eq:NSLeqs} can be written as 
\be
\label{eq:gensolhydrofluc}
\begin{split}
\psi^a(x^0,\bk)\,&=\,\mathcal{U}^{ab}(x^0,0,\bk)\,\psi^b(0)\,\,\\
&+\,\int_{0}^{x^0}\,dt'\,\mathcal{U}^{ab}(x^0,t',\bk)\,\xi^b(t',\bk)\,,
\end{split}
\ee
where $\psi^b (0)$ is the initial condition for the hydrodynamic fluctuating fields at the initial time $x^0=0$ and the matrix operator $\mathcal{U}(t_a,t_b,\bk)$ (with $t_a>t_b$) is defined by
\be
\label{eq:Uoperator}
\mathcal{U}(t_f,t_i,\bk)=\,e^{-\int_{t_i}^{t_f}\,dt'\,\left[\mc A(t',\bk)+\mc D(t',\bk)-\mc P(t',\bk)\right]}\,.
\ee
This operator is a solution of the  differential equation $\partial_0\mc U+ \left(\mc A(t',\bk)+\mc D(t',\bk)-\mc P(t',\bk)\right)\mc U=0$ . The discussion of the mathematical properties of the evolution operators~\eqref{eq:Uoperator} and their application to stochastic processes in statistical mechanics can be found in Refs.~\cite{Kampen,fox1978gaussian}. 

If one performs a stochastic average over the solution~\eqref{eq:gensolhydrofluc} in the absence of external sources then one finds that $\langle\,\psi^a(x^0,\bk)\,\rangle=0$. If instead one considers an ensemble average over different initial conditions then $\langle\langle\,\,\psi^a(x^0,\bk)\,\,\rangle\rangle=\langle\langle\,\,\mathcal{U}^{ab}(x^0)\,\psi^b(0)\,\,\rangle\rangle$ describes the hydrodynamic response to initial state fluctuations. This is an interesting problem in its own right~\cite{Floerchinger:2015efa,Floerchinger:2014fta,Floerchinger:2013tya,Floerchinger:2013hza,Floerchinger:2013vua,Floerchinger:2013rya,Florchinger:2011qf,Kurkela:2018wud,Kurkela:2018vqr,Keegan:2016cpi}, but not the main focus of the present study. 

Information on hydrodynamic fluctuations is contained in two-point and higher n-point correlation functions. We define the symmetric two point correlation function at equal times as~\footnote{In an effective hydrodynamic theory where the fields are purely classical there is no distinction between the symmetrized correlator $\langle\frac{1}{2}\{B(t),A(0)\}\rangle$ and the Wightman correlator $\langle\,B(t)A(0)\,\rangle$~\cite{Kovtun:2003vj}.}
\be 
\label{eq:corrfunc}
\mathcal{C}_{ab}(x^0,\bk)\,\delta^{(3)}(\bk-\bk')
  \equiv\,\Bigl\langle\,\frac{1}{2}\left\{\psi_a(x^0,\bk)\,,\,\psi^\dagger_b(x^0,\bk')\,\right\}\Bigl\rangle\,.
\ee
The reality constraint of the hydrodynamic fluctuating fields implies $\phi^*(x^0,\bk)=\phi(x^0,-\bk)$ so that $C_{ab}$ is a real symmetric matrix. Using the equation of motion for the hydrodynamic variables $\psi_a$, Eq.~\eqref{eq:NSLeqs}, we can derive an evolution equation for the two-point function. We find
\be
\label{eq:corrmat}
\partial_0 \mathcal{C}\,+\,\left[\,\mc A\,,\,\mc C\right]\,+\,\{\,\mc D\,,\,\mc C\,\}\,=\,
\mc P\,\mc C\,+\,\mc C\,\mc P^\dagger\,+\,\frac{1}{2}\left(
\,\mathcal{N}\,+\,\mc N^\dagger\right)\,,
\ee
where $\mathcal{N}$ denotes the two-point correlation function of the noise
\be
\label{eq:xicorr}
\begin{split}
\langle\,\xi_{a,\bk}(x^0_1)\,\xi^{\dagger}_{b,\bk'}(x^0_2)\,\rangle\,&\equiv \mathcal{N}_{ab}(\bk)\\
&\hspace{-2.5cm}=\,2\,(2\pi)^3\,T\,\Hb\,\delta(x^0_1-x^0_2)\,\delta^{(3)}(\bk-\bk')\,\chi_{ac}\,\mc D_{cb}\,. 
\end{split}
\ee
The noise correlator was originally defined in Eqs.~\eqref{eq:noisevar}. Here, we have expressed $\mathcal{N}$ in terms of the diffusive matrix $\mc D_{ab}$ and the susceptibility matrix $\chi_{ab}$. We define $\chi_{ab}$ in Appendix~\ref{app:thermo}.

The equation of motion for the correlation functions $\mc C_{ab}$ mixes different channels. Following the approach of Akamatsu et. al.~\cite{Akamatsu:2016llw} we write the ODE's~\eqref{eq:corrmat} in a basis spanned by the eigenvectors of the acoustic matrix $\mc A$. The acoustic matrix $\mc A$~\eqref{eq:accoustic} is a $5\times 5$ antihermitian matrix (i.e. $\mc A^\dagger=-\mc A$) and the solutions to its eigenproblem $\mc A\,\phi_i\,=\lambda_i\,\phi_i$ define a 5-dimensional complex vector space $\mc H_5$. The eigenvalues of $\mc A$ are purely imaginary or zero. In our case these are given by 
\be
\label{eq:Aeigenval}
\eva=\begin{cases}
i\,v_a &\quad \text{for } i=+\,,\\
-i\,v_a&\quad \text{for } i=-\,,\\
0 &\quad \text{for } i=d,\bold{T}_1,\bold{T}_2\,.
\end{cases}
\ee
In general, the eigenvectors of a non-hermitian matrix need not be mutually orthogonal, and it is necessary to construct a set of left and right eigenvectors. Nonetheless, the antihermitian property of $\mc A$ ensures that the eigenvectors are orthogonal. The eigenvectors $\phi_i$ of $\mc A$~\eqref{eq:accoustic} are given by
\bs
\label{eq:Aeigenvec}
\beal
\phi_\pm&=\frac{1}{\sqrt{2}}\left(1,\pm\hbk,0\right)\,,\\
\phi_d&=\,\left(0,\bold{0},1\right)\,,\\
\phi_{\bold{T}_i}&=\,\left(0,\hat{e}_{\bold{T}_{i}},0\right)\,\hspace{0.5cm}\text{with}\,\,i=1,2,
\end{align}
\es
where $\hbk$ and $\hat{e}_{\bm{T}_l}$ are given by Eqs.~\eqref{eq:3d-spbasis}. In the eigenmode basis Eqs.~\eqref{eq:corrmat} take the form
\be
\label{eq:correigen}
\begin{split}
\partial_0\,\mc C_{AB}\,&+\,k\,\delta\lambda_{AB}\,\mc C_{AB}\,+\,k^2\,\left(\mc D_{AA}\,\mc C_{AB}+\mc C_{AB}\,\mc D_{BB}\right)=\\
&\,\sum_{C}\left(\mc P^\dagger_{AC}\,\mc C_{CB}\,+\,\mc C_{AC}\mc P_{CB}\right)\,+\,\frac{1}{2}\left(\mc N_{AB}+\mc N^\dagger_{BA}\right)\,,
\end{split}
\ee
where $A,B=\{+,-,d,\bold{T}_1,\bold{T}_2\}$. We have also defined $\delta\lambda_{AB} = \lambda_A-\lambda_B$, where $\lambda_{A,B}$ are the eigenvalues of the acoustic matrix $\mc A$. In components the equations of motion can be written as
\bs
\label{eq:evoleqsCAB}
\beal
\label{eq:evoleqsC++}
&\partial_0 \C_{\pm\pm}\,+\,\frac{4}{3}\,\geta\,k^2\,\C_{\pm\pm}=\,\frac{4}{3}\geta\,k^2\,\C_0
\,\nonumber\\
&-\left[\frac{1}{2}\left(\sin^2\theta-2\cos^2\theta\right)\partial_0 h_S\,\mp\,\frac{\hbk_j\,E_j}{ \bw\,v_a}\left(1- v_a^2\right)\right]\,\C_{\pm\pm}\,,\\
\label{eq:evoleqsCTT}
&\partial_0\C_{\bm{T}_1\bm{T}_1}\,+\,2\geta\,k^2\C_{\bm{T}_1\bm{T}_1}\,=\,2\geta\,k^2\C_0
-\,\C_{\bm{T}_1\bm{T}_1}\,\partial_0 h_S\,,\\
&\partial_0\C_{\bm{T}_2\bm{T}_2}\,+\,2\geta\,k^2\C_{\bm{T}_2\bm{T}_2}\,=\,2\geta\,k^2\C_0\,\nonumber\\
&\hspace{1cm}-\C_{\bm{T}_2\bm{T}_2}\,\left(\cos^2\theta-2\sin^2\theta\right)\partial_0 h_S\,,\\
\label{eq:evoleqsCdd}
&\partial_0 \C_{dd}\,+\,2D_0\,k^2\,\C_{dd}=\,2D_0\,k^2\,\chi_{dd}\,\C_0\,,\\
\label{eq:evoleqsCdT-1}
&\partial_0 \C_{d\bm{T}_1}\,+\,\left(D_0+\geta\right)k^2\,\C_{d\bm{T}_1}=
\,-\,\mc C_{d\bm{T}_1}\,\partial_0 h_S\,\nonumber\\
&\hspace{1.5cm}-\frac{1}{\bw}\left(\etr 1\right)_j\,E_j\,\left(\mc C_{\bm{T}_1\bm{T}_1}+\,\mc C_{dd}\right)\,,\\
\label{eq:evoleqsCdT-2}
&\partial_0 \C_{d\bm{T}_2}\,+\,\left(D_0+\geta\right)k^2\,\C_{d\bm{T}_2}=\,-
\frac{1}{\bw}\left(\etr 2\right)_j\,E_j\,\left(\mc C_{\bm{T}_2\bm{T}_2}\,+\,\mc C_{dd}\right)\,\nonumber\\
&\hspace{1.5cm}
\,-\,\mc C_{d\bm{T}_2}\,\left(\cos^2\theta-2\sin^2\theta\right)\partial_0 h_S\,.
\end{align}
\es
with $\C_0=\bar{T}\,\Hb$. In a recent work Akamatsu et. al.~\cite{Akamatsu:2018vjr} studied the dynamics of  Eqs.~\eqref{eq:evoleqsC++}-\eqref{eq:evoleqsCdd} for a generic EOS. 
For the purpose of determining hydrodynamic tails we are interested in the long-time, low-frequency behavior of the correlation functions. In the long-time limit diagonal correlation functions relax to equilibrium states fixed by fluctuation-dissipation relations. External sources lead to small deviations that can be studied perturbatively. The solutions of Eqs.~\eqref{eq:evoleqsC++}-\eqref{eq:evoleqsCTT} are given by
\be
\label{eq:asymptsols}
\mc C_{AA}(\omega,\bk)=\,T\,\Hb\,\chi_{AA}\left((2\pi)\delta (\omega)\,+\,\frac{\mc P_{AA}(\omega,\bk)}{-i\omega\,+\,2\,k^2\,\mc D_{AA}}
\right)\,,
\ee
where 
\be
\begin{split}
&\mc P_{\pm\pm}= i\omega\left[h(\omega)\left(\sin^2\theta-2\cos^2\theta\right)\,\pm\,\frac{\hbk_i\,\delta A_i(\omega)}{\bw\,v_a}\left(1-v_a^2\right)\right]
\,,\\
&\mc P_{\bm{T}_1\bm{T}_1}=i\omega\,h(\omega)\,,\hspace{.5cm}\mc P_{\bm{T}_2\bm{T}_2}=\,i\omega\,\left(\cos^2\theta-2\sin^2\theta\right)h(\omega)\,,\\
&\mc D_{\pm\pm}= \frac{2}{3}\geta\,,\hspace{.5cm}
\mc D_{\bm{T}_l\bm{T}_l}=\geta\,,\hspace{.5cm}\mc D_{dd}=\,D_0\,.
\end{split}
\ee
The correlation function $\C_{dd}$ does not directly couple to a source and is given by
\be
\label{eq:asymptsolCdd}
\C_{dd}=\chi_{dd}\,\C_0\,.
\ee
Off-diagonal correlation functions relax to zero, but external sources, coupled to the equilibrium behavior of $C_{\bm{T}_l\bm{T}_l}$ and $C_{dd}$, can drive them off equilibrium. The asymptotic behavior of $\C_{d\bm{T}_l}$ is 
\be
\label{eq:asymptsolCdt}
\C_{d\bm{T}_l}=\,-i\omega\,\left[\frac{\left(\hat{e}_{\bm{T}_l}\right)_j\,\delta A_j(\omega,\bk)}{-i\omega+\left(D_0+\geta\right)k^2}\right]\,\frac{T\,\Hb}{\bw}\,\chi_{d\bm{T}}\,,
\ee
where we have defined $\chi_{d\bm{T}_l}=\chi_{\bm{T}_l\bm{T}_l}+\chi_{dd}$.

\section{Modification of the response functions due to hydrodynamic fluctuations}
\label{sec:mod}

For a scale-invariant theory and in the limit when ${\bm k}\to 0$ (with $w$ fixed) the response functions of the energy-momentum tensor and the charge current are defined by
\bs
\label{eq:GF}
\beal
\label{eq:GFtmn}
\delta \tmn(\omega)&=-\frac{1}{2}\,\lim_{{\bm k}\to 0} G^{\mu\nu,\lambda\sigma}_R(\omega,\bk)\,h_{\lambda\sigma}(\omega,\bk)\,,\\
\label{eq:GFcurrent}
\delta J^\mu(\omega)&=-\lim_{{\bm k}\to 0} \,G^{\mu\nu}_R(\omega,\bk)\,\delta A_\nu(\omega,\bk)\,,
\end{align}
\es
In this limit the response functions $G^{\mu\nu,\lambda\sigma}_R$ and $G^{\mu\nu}_R$ can be written as~\cite{Czajka:2017bod,Kovtun:2005ev,yaffe}
\bs
\label{eq:GFdecomp}
\beal
G^{\mu\nu,\lambda\sigma}_R&= \,H^{\mu\nu,\lambda\sigma}\,G_S^R\,,\\
G^{\mu\nu}_R&= P^{\mu\nu}G_J^R\,.
\end{align}
\es
$G_S^R$ and $G_J^R$ are scalar functions while $P^{\mu\nu}$ and $H^{\mu\nu,\lambda\sigma}$ (for d=4) are given by
\bs
\label{eq:proj}
\beal
P^{\mu\nu}&=\gup-\frac{k^\mu\,k^\nu}{k^\mu\,k_\mu}\,,\\
\label{eq:Hop}
H^{\mu\nu,\lambda\sigma}&=\frac{1}{2}\left(P^{\mu\lambda}P^{\nu\sigma}+P^{\mu\sigma}P^{\nu\lambda}\right)-\frac{1}{3}P^{\mu\nu}\,P^{\lambda\sigma}\,\,.
\end{align}
\es
The operator $H^{\mu\nu,\lambda\sigma}$ is a projector onto conserved traceless symmetric tensors which satisfies $\eta_{\mu\nu}\,H^{\mu\nu,\lambda\sigma} = 0$~\cite{Czajka:2017bod,Kovtun:2005ev,yaffe}. In the limit $\bk\to 0$ the scalar functions $G_S$ and $G_J$ can be written in terms of $\delta T^{ij}$ and $\delta J^i$ as follows 
\bs
\label{eq:GFsols}
\beal
\label{eq:GS}
G^R_S(\omega)&=-\frac{1}{6}\,\lim_{{\bm k}\to 0}\,\left(\frac{\delta T^{11}(\omega,\bk)}{\delta h_s(\omega,\bk)}+\frac{\delta T^{22}(\omega,\bk)}{\delta h_s(\omega,\bk)}\right.\nonumber\\
&\left.-2\,\frac{\delta T^{33}(\omega,\bk)}{\delta h_s(\omega,\bk)}\,\right),\\
\label{eq:GJ}
G_J^R(\omega)&=-\lim_{{\bm k}\to 0}\,\delta^j_{\hspace{.1cm}i}\frac{\delta J^i(\omega,\bk)}{\delta A^j(\omega,\bk)}\,.
\end{align}
\es
In hydrodynamics we can read off the variation of the spatial components of the energy-momentum tensor~\eqref{eq:tmndecom} and the charge current~\eqref{eq:jmdecom} 
from the constitutive relations. Taking into account both the effects of external fields and the variations of the hydrodynamic variable we obtain
\bs
\label{eq:vartmnj}
\beal
\label{eq:vartmn}
\delta T^{ij}\,&=\,\left(\,\pro\,+\,\dep\,\right)\dij\,+\,\pro\,\hij\,+\,\frac{H^{ij}_{kl}}{\Hb}\,\frg^k\,\frg^l\,\nonumber\\
&-\geta\,\left(\partial^i\frg^j+\partial^j\frg^i-\frac{2}{3}\delta^{ij}\partial_k\frg_k\right)\,-\,\geta\,\Hb\partial_0 h^{ij}\,,\\
\label{eq:varj}
\delta J^i&=\,\frac{\frg^i}{\bar{w}}\,+\,\frac{1}{\Hb\,\bar{w}}\,\left[\frac{1}{v_a}\left(\frac{\dep}{v_a}\right)-\dq\right]\,\frg^i\,\nonumber\\
&+\,\sigma_0\,F^{i0}\,+\,\frac{D}{\bar{w}}\,\partial^i\dq\,.
\end{align}
\es
where we include corrections up to second order in fluctuations. These terms describe the non-linear mode-coupling between the currents and the hydrodynamic variables. Note that the second order terms are determined by the symmetric two-point functions~\eqref{eq:corrfunc} studied in the previous section~\cite{Kovtun:2012rj,Kovtun:2003vj,Akamatsu:2016llw,pomeau1975time,fox1978gaussian}
\be
\label{eq:bracorr}
\begin{split}
\langle\,\Psi_a (x^0,\bm{x})\,\Psi_b^\dagger (x^0,\bm{x'})\,\rangle\,&=\,\int_{\bk}\,\C_{ab}(x^0,\bm{k})\,,\\
&\hspace{-1.5cm}=\,\sum_{I,J}\,\int_{\bk}\,\left(e_A\right)_a\,\tilde{\C}_{AB}(x^0,\bm{k})\,\left(e_B\right)_b\,,
\end{split}
\ee
Here, we have expressed the symmetric correlation matrix $\C_{ab}$~\eqref{eq:corrfunc} in terms of the eigenmodes of the acoustic matrix $\mc A$. This provides a transparent interpretation of different contribution to the currents in terms of hydrodynamic modes~\cite{landauv9,kadanoffmartin,forster1995hydrodynamic,Chafin:2012eq,Kovtun:2003vj,Akamatsu:2016llw,Kovtun:2011np,Martinez:2017jjf}. 

\section{Long time tails: Static case}
\label{sec:static}

As a first application of this formalism we compute the low frequency behavior of the response in the static case. For harmonic perturbations $h_s, \delta A \sim e^{i\omega x^0}$ Eqs.~\eqref{eq:vartmnj} give
\bs
\label{eq:confvartmnj}
\beal
\label{eq:confvartmn}
\langle\,\delta T^{ij}(\omega)\,\rangle\,&=\,\left(\,\pro\,+\,\dep\,\right)\dij\,+\,\pro\,\hij\,\nonumber\\
&+\,\frac{H^{ij}_{kl}}{\Hb}\,g^{kr}\,g^{ls}\,\langle\,\frg_r\,\frg_s\,\rangle\,-\,i\omega\,\geta\,\Hb\,h^{ij}\,,\\
\label{eq:confvarj}
\langle\,\delta J^i(\omega)\,\rangle\,&=\,\frac{\frg^j}{\bar{w}}\,+\,\frac{1}{\Hb\,\bar{w}}\,\left[\frac{1}{v_a}\langle\,\left(\dep/v_a\,\right)\,\frg^j\,\rangle-\langle\,\dq\,\,\frg^j\rangle\right]\,\,\nonumber\\
&+\,i\,\omega\,\sigma_0\,\delta A\,\hbk^i\,.
\end{align}
\es
In Eq.~\eqref{eq:confvartmnj} the non-linear contributions of the hydrodynamic fluctuations are encoded in the correlators $\langle\,\frg_i\,\frg_j\,\rangle$, $\langle\left(\dep/v_a\right)\,\frg_i\,\rangle$ and $\langle\,\dq\,\frg_i\,\rangle$. In terms of the eigenmodes of the acoustic matrix~\eqref{eq:Aeigenvec} these correlators are
given by
\bs
\label{eq:leadingcorr}
\beal
\label{eq:leadingcorr-gg}
\langle\,\frg_i\,\frg_j\,\rangle\,&\equiv\,\C _{\frg_i\,\frg_j}(\omega,\bk)\nonumber\\
&\approx\frac{1}{4}\hbk_i\hbk_j\,\left[ \tilde{\C}_{++}(\omega,\bk)+\tilde{\C}_{--}(\omega,\bk)\right]\,\nonumber\\
&+\,\sum_{l=1}^2\,\tilde{\C}_{\bm{T}_l\bm{T}_l}(\omega,\bk)\,\left(\etr l \right)_i\left(\etr l \right)_j\,,\\
\label{eq:leadingcorr-pg}
\left\langle\,\left(\frac{\dep}{v_a}\right)\frg_i\,\right\rangle\,&\equiv\,\C _{\dep\,\frg_i}(\omega,\bk)\nonumber\\
&\approx\,\frac{1}{4}\left[ \tilde{\C}_{++}(\omega,\bk)-\tilde{\C}_{--}(\omega,\bk)\right]\hbk_i\,,\\
\label{eq:leadingcorr-qg}
\left\langle\,\dq\,\frg_i\,\right\rangle\,&\equiv\,\C _{\dq\,\frg_i}(\omega,\bk)\nonumber\\
&\approx\,\sum_{l=1}^2\,\tilde{\C}_{d\bm{T}_l}(\omega,\bk)\,\left(\etr l \right)_i
\end{align} 
\es
Here, we have neglected terms that correspond to rapidly oscillating modes~\cite{Akamatsu:2016llw,Akamatsu:2017rdu}. In the language of the diagrammatic approach these are diagrams that have poles far away from the real axis in frequency space~\cite{Kovtun:2003vj,Kovtun:2011np,Martinez:2017jjf,Chafin:2012eq}. 

We observe that the leading term in the stress tensor response, Eq.~\eqref{eq:confvartmn} and ~\eqref{eq:leadingcorr-gg}, only contains diagonal correlation functions~\cite{Akamatsu:2016llw,Akamatsu:2017rdu}. The current response, on the other hand, is sensitive to off-diagonal correlation functions, see Eq.~\eqref{eq:confvarj} and \eqref{eq:leadingcorr-qg}. This is interesting, because in the diagrammatic approach the current response function is factorized into diagonal propagators, see Kovtun and Yaffe~\cite{Kovtun:2003vj} as well as App.~\ref{app:supp}. We will show that the result (for the modes taken into account in both works) nevertheless agrees. 

We have all the ingredients to calculate the response functions~\eqref{eq:GFsols}. The stress tensor response $G_S^R$~\eqref{eq:GS} was obtained by Akamatsu et. al.~\cite{Akamatsu:2016llw} and we simply state their result
\begin{widetext}
\be
\label{eq:corrtmn}
\begin{split}
G_{S}^R(\omega)&=\pro-i\omega \Hb\geta-\frac{T}{6}\int_{\bold k}\left[-6\,+\,i\omega\,\frac{\left(\sin^2\theta-2\cos^2\theta\right)^2}{-i\omega+\frac{4}{3}\geta k^2}\,+\,i\omega\frac{\left(1+(\cos^2\theta-2\sin^2\theta)^2\right)}{-i\omega+2\geta k^2}
\right]	\,,\\
&=\pro+\frac{T\,\Lambda^3}{6\pi^2}-i\omega\left(\eta_0+\frac{\Lambda}{\geta}\frac{17}{120\pi^2}T\right)\,+\,\frac{(1+i)}{\geta^{3/2}}\frac{\left[\left(\frac{3}{2}\right)^{3/2}+7\right]}{240\pi}\,T\omega^{3/2}\,,\\
\end{split}
\ee
\end{widetext}
where we have used $v_a^2=1/3$ as well as $\chi_{AA}=2$ for $A=\pm$ and $\chi_{AA}=1$ for $A=\bm{T}_l$. In Eq.~\eqref{eq:corrtmn} we have regularized UV divergent integrals using a sharp cutoff $\Lambda$ in momentum space. There are two divergent terms, a $\mc O(\Lambda^3)$ term that can be viewed as a correction to the pressure, and a $\mc O(\Lambda)$ contribution that modifies the shear viscosity. Stochastic hydrodynamics is renormalizable in the sense that divergent terms can be absorbed in the parameters of the theory, the equation of state and the transport coefficients. The main result is the leading non-analytic term $\mc O(\omega^{3/2})$, which is independent of the cutoff. This term corresponds to a $t^{-3/2}$ decay of the real time correlation function and is known as a hydrodynamic tail. The tail of the relativistic stress tensor correlation function was computed in \cite{Kovtun:2003vj,Kovtun:2011np}, and checked using the formalism employed in this work in~\cite{Akamatsu:2016llw}.

Using the same methods we can determine the current response function $G_J^R$~\eqref{eq:GJ}. We find
\begin{widetext}
\be
\label{eq:corrJJ}
\begin{split}
G_J^R(\omega)&=\,i\omega\,\sigma_0\,\delta_{ij}+\,i\omega\,\left(\frac{T}{\bw^2}\right)\,\left[
\int_{\bk}\,\frac{\hbk_i\hbk_j}{-i\omega+\frac{4}{3}\geta\,k^2}\,+\,\sum_{l=1}^2\,\chi_{d\bm{T}_l}\,\int_{\bk}\,\frac{\left(\hat{e}_{\bm{T}_l}\right)_i\,\left(\hat{e}_{\bm{T}_l}\right)_j}{-i\omega+(D+\geta)k^2}
\right]\,,\\
&=\,i\,\delta_{ij}\,\omega\,\left[\sigma_0+\,\frac{T}{\bw^2}\,\frac{\Lambda}{\pi^2}\left(\frac{1}{8\,\geta}+\frac{T\,\,c_p}{3\,\Hb\,(D+\geta)}\right)\right]\\
&\hspace{1cm}+\,\frac{(1+i)}{\sqrt{2}}\frac{\omega^{3/2}}{\pi}\left(\frac{1}{12}\left(\frac{3}{4}\right)^{3/2}\frac{1}{\geta^{3/2}}\,+\,\frac{T\,c_p}{6\,\Hb}\frac{1}{(D+\geta)^{3/2}}\right)\,,\\
\end{split}
\ee
\end{widetext}
where we have approximated $\chi_{d\bm{T_l}}=T\,c_p/\Hb$, where $c_p$ is the specific heat at constant pressure, to facilitate the comparison with diagrammatic calculations, see App.~\ref{app:supp}. Also see App.~\ref{app:thermo} for the definition of thermodynamic quantities. We note that the current response does not contain a cubic divergence. There is a linear divergence which can be absorbed into the conductivity. We find a non-analytic contribution, which is again of the form $\omega^{3/2}$. It receives contributions from sound modes, proportional to $\gamma_{\eta_0}^{-3/2}$, and from the mixing of diffusive heat and shear waves, governed by $(D+\gamma_{\eta_0})^{-3/2}$. Note that in a conformal fluid sound attenuation is purely governed by $\gamma_{\eta_0}$. The purely diffusive terms were previously computed by Kovtun and Yaffe~\cite{Kovtun:2003vj}~\footnote{It is important to mention that in Sect. III of Ref.~\cite{Kovtun:2003vj},  Kovtun and Yaffe studied the $N$=1 supersymmetric hydrodynamic theory in $d=4$. The $\langle J_R J_R\,\rangle$ symmetric correlator for the conserved $R$-charge of this supersymmetric theory (Eq. 83) is similar to our result~\eqref{eq:corrJJ}. Nonetheless, this supersymmetric field theory has more degrees of freedom than the single $U(1)$ charge fluid studied here and thus, different couplings arise at the level of the constitutive relations.}. We provide a more detailed comparison with their work in App.~\ref{app:supp}.

\subsubsection{Spectral functions}
\label{sec:renormalization}

The imaginary part of the response function determines the spectral function, and it can be used to define frequency dependent diffusion constants. We find
\bs
\label{eq:renor-w}
\beal
\label{eq:renshear-w}
\gamma_{\eta}(\omega)&=-\frac{\text{Im}\left[G^R_S(\omega,\bk=0)\right]}{\Hb\,\omega}\,,\nonumber\\
&=\gamma_\eta^{ren.}\,-\,\frac{T}{\Hb \gamma_\eta^{ren.}}\frac{\left[\left(\frac{3}{2}\right)^{3/2}+7\right]}{240\pi}\,\omega^{1/2}\,,\\
\label{eq:rencond-w}
D(\omega)&=-\alpha_2\,\frac{\text{Im}\left[G^R_J(\omega,\bk=0)\right]}{\omega}\,,\nonumber\\
&=D^{ren.}\,-\,\frac{\omega^{1/2}}{\sqrt{2}\pi}\,\alpha_2\,\left[\frac{1}{12}\left(\frac{3}{4}\right)^{3/2}\frac{1}{[\gamma_\eta^{ren.}]^{3/2}}\,\right.\nonumber\\
&\hspace{2cm}\left.+\,\frac{T\,c_p}{6\,\Hb}\frac{1}{(D^{ren.}+\gamma^{ren.}_\eta)^{3/2}}\right]\,.
\end{align}
\es
where the renormalized momentum diffusion constant $\gamma_{\eta}^{ren.}$ and renormalized charge diffusion constant $D^{ren.}$ are given by
\bs
\label{eq:renor}
\beal
\label{eq:renor-gam}
\gamma_\eta^{ren.}&=\geta+\frac{17}{120\pi^2}\frac{T}{\Hb}\,\frac{\Lambda}{\gamma_{\eta}^{ren.}} \,,\\
\label{eq:renor-D}
D^{ren.}&= D_0+\alpha_2\frac{T}{\bw^2}\,\frac{\Lambda}{\pi^2}\left(\frac{1}{8\,\gamma_\eta^{ren.}}\right.\nonumber\\
&\left.+\frac{T\,c_p}{3\,\Hb\,(D^{ren.}+\gamma_\eta^{ren.})}\right)\,.
\end{align}
\es
We note that the spectral function has a $\omega^{1/2}$ non-analyticity near $\omega=0$, and that the spectral functions are enhanced in the small frequency limit. We also note that the negative sign of the $\omega^{1/2}$ terms implies that the zero-frequency diffusion constants cannot be arbitrarily small. 

\subsubsection{Lower bounds of the transport coefficients}
\label{sec:lower}

The renormalized transport coefficients, Eqs.~\eqref{eq:renor}, are given as the sum of a bare contribution and a loop correction which scales inversely with the bare coupling. This implies that the renormalized transport coefficients cannot be arbitrarily small. The physical mechanism underlying this bound is the fact that even if the microscopic diffusion constant is small there is always a macroscopic mechanism, associated with the propagation of hydrodynamic modes, that leads to diffusion. 

More formally, we can view this bound as arising from a matching procedure. Consider a UV scale $\Lambda$ at which the long wavelenght hydrodynamic theory is matched to a microscopic theory with transport coefficients $\gamma_{\eta_0}$ and $D_0$. Consistency requires that both $\gamma_{\eta_0}$ and $D_0$ are positive. Contributions to the transport coefficients from scales below $\Lambda$ can be computed in stochastic hydrodynamics, and lead to terms that scale with a positive power of the cutoff, and inversely with the bare transport coefficient. This means that for any value of the UV scale there is a bound on the transport coefficient, and that the bound is stronger the larger the cutoff can be chosen. 

The largest possible cutoff scale is determined by the regime of validity of fluid dynamics. Consider a diffusive shear mode. The decay rate is proportional to $\omega\sim \gamma_\eta k^2$. For hydrodynamics to be valid this rate has to be smaller than the next order gradient correction, the shear relaxation rate $\tau_\pi^{-1}$. This means $\gamma_\eta k^2<\tau_\pi^{-1}$ and $\Lambda=(\gamma_\eta\,\tau_\pi)^{-1/2}$ . In microscopic theories, both in the weak and strong coupling limit, the diffusion constant and the relaxation time are driven by similar physical processes, and $\tau_\pi=\gamma_\eta  \lambda(p(T,\mu))$~\cite{Moore:2012tc,Moore:2010bu,York:2008rr,Baier:2007ix,Bhattacharyya:2008jc,Erdmenger:2008rm,Banerjee:2008th}, where $\lambda$ is a function of the thermodynamic pressure $p$. We conclude that 
\be 
\label{hydro-cut}
\Lambda_{\it max} = \frac{1}{\gamma_\eta \lambda^{1/2}}\, . 
\ee
Similar estimates can be obtained by considering the sound and heat diffusion channels~\cite{Kovtun:2012rj,Martinez:2017jjf}, and we will assume that Eq.~\eqref{hydro-cut} can be applied in all channels. 

For $\Lambda=\Lambda_{max.}$ the physical values of the sound attenuation length and the diffusion coefficient, Eqs.~\eqref{eq:renor} read as
\bs
\beal
\label{eq:maxrenor-gam}
\gamma_\eta^{ren.}(0)\left.\right|_{\Lambda_{max}}&\nonumber\\
&\hspace{-1.5cm}=\geta +\frac{17}{120\pi^2}\frac{T}{\Hb}\,\frac{1}{[\gamma_{\eta}]^{2}\,\lambda^{1/2}}\,,\\
\label{eq:maxrenor-D}
D^{ren.}(0)\left.\right|_{\Lambda_{max}}&\nonumber\\
&\hspace{-1.5cm}= D_0+\frac{\alpha_2}{\pi^2}\frac{T}{\bw^2}\frac{1}{\gamma_\eta\,\lambda^{1/2}}\left(\frac{1}{8\,\gamma_\eta}\right.\nonumber\\
&\left.+\frac{T\,c_p}{3\,\Hb\,(D+\gamma_\eta)}\right)\,.
\end{align}
\es
By extremizing the first of these expressions respect to $\gamma_\eta$ we get the following lower bound for $\gamma_\eta$
\be
\label{eq:lowgeta}
\gamma_\eta \geq \,\gamma_\eta^{min.}=0.4593\left[\frac{T}{\Hb}\,\frac{1}{\lambda^{1/2}}\right]^{1/3}\,.
\ee
This result was found previously by Kovtun et. al.~\cite{Kovtun:2011np} for vanishing chemical potential so $\lambda\equiv\lambda(p(T))$ in Eq.~(4.3) of Ref.~\cite{Kovtun:2011np}. Now the lower bound of the diffusion coefficient is obtained by considering the bare diffusion coefficient to vanish exactly $D_0=0$ in Eq.~\eqref{eq:maxrenor-D} such that 
\be
\label{eq:lowD}
\begin{split}
D&\geq D\left.\right|_{\geta}\,,\\
&=
\left[\frac{\alpha_2}{\pi^2}\frac{T}{\bw^2}\frac{1}{\geta^2\,\lambda^{1/2}}\left(\frac{1}{8}
+\frac{T\,c_p}{3\,\Hb}\right)\right]\equiv \frac{\Xi}{\geta^2}
\,.
\end{split}
\ee  
Therefore, the lower bound of the diffusion coefficient depend exclusively on $\geta$. The lower bounds of the attenuation transport coefficients also depend on the choice of the equation of state. 

\section{Long time tails: Bjorken expansion case}
\label{sec:Bjorken}

We now come to the central problem studied in this work, extending the calculation of diffusive tails to an expanding background. We will consider a conformal U(1) charged fluid undergoing longitudinal boost invariant expansion, known as Bjorken flow. The symmetries of the Bjorken flow  are manifest in Milne coordinates 
\be
x^\mu=(\tau,\bx_\perp,\varsigma), 
\ee
where $\tau=\sqrt{(x^0)^2-(x^3)^2}$ is the proper time, $\varsigma = \, {\rm arctanh} (x^3/x^0)$ is spatial rapidity, and the metric is $g_{\mu\nu} = \,{\rm diag}.~(-1,1,1,\tau^2)$. 

In the absence of external sources and noise the conservation laws provide the evolution equations for the background hydrodynamic fields. Bjorken flow corresponds to the well known equations
\bs
\label{eq:backEOM}
\beal
\partial_\tau\ene\,+\,\frac{\ene+\pro}{\tau}&=0\,,\\
\partial_\tau\bd\,+\,\frac{\bd}{\tau}&=0\,,
\end{align}
\es
where we have neglected dissipative terms. For the conformal EOS~\eqref{eq:confeos} the hydrodynamic background fields evolve according to 
\be 
\ene(\tau)=\epsilon_0\,\left(\frac{\tau_0}{\tau}\right)^{1+v_a^2}\, , 
\hspace{0.5cm}
\bd(\tau)=n_0\,\left(\frac{\tau_0}{\tau}\right)\, . 
\ee
We also find the background pressure $\pro(\tau)=v_a^2\,\ene(\tau)$ as well as the  heat energy density $\bar{q}=\ene(\tau)\:-\:\bw(\tau)\:\bd(\tau)$.  

The evolution equations for the hydrodynamic variables in the presence of external fields and noise can be obtained using the procedure discussed in Sect.~\ref{subsec:L-NS}. We again adopt a mixed representation and write define spatial momenta $\bk\equiv(\underline{\bk},k_\varsigma)$. The spatial Fourier transform is defined as
\be
A(\tau,\bm{k})    
    =\int\,d\varsigma\,d^2\bx\,e^{-i\left(\bx^\perp\cdot\ubk\, + \,\kappa \varsigma\right)} \,A(t,\bx_\perp,\varsigma)\,.
\ee
with $\kappa=k_\varsigma$. The linearized stochastic hydrodynamic equations are~
\begin{widetext}
\bs
\label{eq:explicitBjoeom}
\beal
\label{eq:Bjoener}
&\partial_\tau\den\,+\,i\,\left(\ubk\cdot\ufrg\,+\tau^2k^\varsigma\,\frg^\varsigma\right)=-\frac{(1+v_a^2)}{\tau}\den\,-\,\frac{1}{\bw}\frg^\perp\,E^{\perp}\,-\,\frac{1}{\bw}\tau^2\,\frg^\varsigma E^{\varsigma}\,,\\
\label{eq:Bjomomfluxperp}
&\partial_\tau\ufrg+i\,v_a^2\ubk\den+\geta\left[\left((\ubk)^2+\left(\tau\,k^\varsigma\right)^2\right)\ufrg+\frac{1}{3}\ubk\,\left(\ubk\cdot\ufrg\,+\tau^2k^\varsigma\,\frg^\varsigma\right)\right]=-\frac{\ufrg}{\tau}+\dbd\,E^{\perp}\,-\,i(\bm{k}_\perp)_i\,S^{i\perp}\\
\label{eq:Bjomomfluxlong}
&\partial_\tau\frg^\varsigma+i\,v_a^2k^\varsigma\,\den\,+\,\geta\left[\left((\ubk)^2+\left(\tau\,k^\varsigma\right)^2\right)\frg^\varsigma\,+\,\frac{1}{3}\,k^\varsigma\,\left(\ubk\cdot\ufrg\,+\,\tau^2k^\varsigma\frg^\varsigma\right)\right]=-3\,\frac{\frg^\varsigma}{\tau}+\dbd\,E^{\varsigma}\,-\,i\tau^2\,k^\varsigma\,S^{\varsigma\varsigma}\\
\label{eq:Bjopart}
&\partial_\tau\dbd\,+i\,\frac{\left(\ubk\cdot\ufrg\,+\,\tau^2\,k^\varsigma\frg^\varsigma\right)}{\bw}+ \sigma_0\left((\ubk)^2+\left(\tau\,k^\varsigma\right)^2\right)\left(\alpha_1\den+\alpha_2\dbd\right)=-\frac{\dbd}{\tau}\,-\,i\,\left(\bm{k}_\perp\cdot \bm{I}_\perp+\tau^2\,k^\varsigma\,I^\varsigma\right)\,,
\end{align}
\es
\end{widetext}
where the electric field is $E^i=F^{i\tau}\,u_\tau\equiv\,F^{\tau i}$. We adopt the basis 
\be 
\psi_a(\tau,\bk)=\left(\dep/v_a,\tau\frg^\varsigma,\frg^\perp,\dq\right)
\ee
and write the Navier-Stokes-Langevin equations in the matrix form~\footnote{Here we assume that the background is slowly evolving, so that $\partial_\tau\delta q=\partial_\tau\delta\epsilon-\bar{w}\partial_\tau\delta n$.}
\be
\label{eq:BjoEOM}
\partial_\tau \psi^a_\bk\,+\,\left(\mc A^{ab}(\tau)+\mc D^{ab}(\tau)\right)\psi^b_\bk=\mc P^{ab}(\tau)\psi^b_\bk\,+\,\xi^a\, .
\ee
Here, the acoustic and diffusive matrices are 
\bs
\label{eq:matrixopBjo}
\beal
\mc A&=K\begin{pmatrix}
0&iv_a\hbbk_i&0\\
iv_a\hbbk_i&0&0\\
0&0&0
\end{pmatrix}\,,\\
\mc D&=K^2\,\text{diag.}\,\,\left(0,\geta\left(\delta_{ij}+\frac{1}{3}\hbbk_i\,\hbbk_j\right),D_0\right)\,,
\end{align}
\es
and the source matrix and noise vector are given by
\bs
\label{eq:matrixopBjo2}
\beal
\mc P&=\begin{pmatrix}
-(1+v_a^2)/\tau&-\frac{v_a}{\bw}\mc E^\varsigma&-\frac{v_a}{\bw}\bm{\mc E}^\perp&0\\
\mc E^\varsigma/(\bw v_a)&-2/\tau&\bm{0}&-\mc E^\varsigma/\bw\\
\bm{\mc E^\perp}/(\bw v_a)&0&-\mathbb{1}_2/\tau&-\bm{\mc E^\perp}/\bw\\
0&-\frac{1}{\bw}\mc E^\varsigma&-\frac{1}{\bw}\bm{\mc E}^\perp&-1/\tau
\end{pmatrix}\,,\\
\xi&=\begin{pmatrix}
0\\\
-\,i\kappa\,S^{\varsigma\varsigma}\\
-\,i(\bm{k}_\perp)_i\,S^{i\perp}\\ 
i\,\bw\,\left(\bm{k}_\perp\cdot \bm{I}_\perp+\kappa\,I^\varsigma\right)
\end{pmatrix}\,,
\end{align}
\es
where $\mc E=(\bm{E}^\perp,\tau E^\varsigma)$, $\mathbb{1}_2$ is the identity in two dimensions, $\kappa=\tau^2 k^\varsigma$ and we define 
\bs
\label{eq:Kvector}
\beal
\overrightarrow{\bm{K}}&=\left(\bk_\perp,\frac{\kappa}{\tau}\right)\,,\\
\hbbk&=\frac{\overrightarrow{\bm{K}}}{K}\equiv\left(\sin\theta_{\bbk}\,\cos\varphi_{\bbk},\sin\theta_{\bbk}\sin\varphi_{\bbk},\cos\theta_{\bbk}\right)\,.
\end{align}
\es
Both $\overrightarrow{\bm{K}}$ and $\hbbk$ are time depending vectors~\cite{Akamatsu:2016llw}. Moreover, the $ISO(2)$ symmetry of the background hydrodynamic fields forbids the azimuthal angle $\varphi_{\bbk}$ to be time dependent. 

The acoustic matrix has the same structure as its static counterpart~\eqref{eq:accoustic}, and the set of orthonormal eigenvectors is given by ~\cite{Akamatsu:2016llw}
\bs
\label{eq:eigenBjo}
\beal
\phi_\pm&=\frac{1}{\sqrt{2}}\left(1,\pm\hbbk,0\right)\,,\\
\phi_{\bm{T}_l}&=\left(0,\hat{e}_{\bm{T}_l},0\right)\,,\\
\phi_{d}&=\left(0,\bm{0},1\right)\,,
\end{align}
\es
where we have defined
\bs
\beal
\etr 1&=\left(-\sin\varphi_{\bbk},\cos\varphi_{\bbk},0\right)\,,\\
\etr 2&=\left(\cos\theta_{\bbk}\cos\varphi_{\bbk},\cos\theta_{\bbk}\sin\varphi_{\bbk},-\sin\theta_{\bbk}\right)\,.
\end{align}
\es
The evolution equations for the symmetric correlation functions are determined by following the procedure discussed in Sect.~\ref{sec:sol}. In the following we will make use of the following components of $\mc C_{AB}$ 
\begin{widetext}
\bs
\label{eq:EOMBcorrjo}
\beal
&\partial_\tau\mc C_{\pm\pm}\,+\,\frac{4}{3}\geta\bbk^2\mc C_{\pm\pm}\,=\,\frac{4}{3}\geta\bbk^2\,\frac{\C_{0}(\tau)}{\tau}-\frac{(2+v_a^2+\cos^2\theta_{\bbk})}{\tau}\,\C_{\pm\pm}\,\pm\,\frac{\hbbk\cdot\mc E}{\bw}\frac{(1-v_a^2)}{\,v_a}\,\C_{\pm\pm}\,,\\
&\partial_\tau\C_{\bm{T}_1\bm{T}_1}\,+\,2\geta\bbk^2\C_{\bm{T}_1\bm{T}_1}\,=\,2\geta\bbk^2\,\frac{\tilde{\C}_{0}(\tau)}{\tau}-\frac{2}{\tau}\,\C_{\bm{T}_1\bm{T}_1}\,,
\\
&\partial_\tau\C_{\bm{T}_2\bm{T}_2}\,+\,2\geta\bbk^2\C_{\bm{T}_2\bm{T}_2}\,=\,2
\geta\bbk^2\,\frac{\tilde{\C}_{0}(\tau)}{\tau}-\frac{2\left(1+\sin^2\theta_{\bm K}\right)}{\tau}\,\C_{\bm{T}_2\bm{T}_2}\,,
\\
&\partial_\tau\C_{dd}\,+\,2\,D_0\bbk^2\,\C_{dd}=\,2\,D_0\,\bbk^2\,\frac{\chi_{dd}\,\tilde{\C}_{0}(\tau)}{\tau}\,-\,\frac{2}{\tau}\,\C_{dd}\,,
\\
&\partial_\tau\C_{d\bm{T}_1}+\left(\geta+D_0\right)\bbk^2\,\C_{d\bm{T}_1}\,=\,-\frac{2}{\tau}\C_{d\bm{T}_1}\,-\,\frac{1}{\bw}\,\hat{e}_{\bm{T}_1}\cdot\,\mc E\,\left(\C_{\bm{T}_1\bm{T}_1}\,+\,\C_{dd}\right)\,,\\
&\partial_\tau\C_{d\bm{T}_2}+\left(\geta+D_0\right)\bbk^2\,\C_{d\bm{T}_2}\,=\,-\frac{2(1+\sin^2\theta_{\bm K})}{\tau}\C_{d\bm{T}_2}\,-\,\frac{1}{\bw}\,\hat{e}_{\bm{T}_2}\cdot\,\mc E\,\left(\C_{\bm{T}_2\bm{T}_2}\,+\,\C_{dd}\right)\,,
\end{align}
\es
\end{widetext}
with $\tilde{\C}_0=T(\tau)\Hb(\tau)$. We have also used the noise correlator 
\be
\begin{split}
\label{eq:noBjo}
\langle\,\xi^a(\tau,\bm{K},\xi^b(\tau',-\bm{K}))\,\rangle&=\,\frac{2\,T\,\Hb}{\tau}\,\chi_{ac}\,\tilde{\mc D}_{cb}\,\\
&\hspace{-2.5cm}\left(2\pi\right)^3\,\delta^{(2)}\left(\bk-\bk'\right)\,\left(\kappa-\kappa'\right)\delta(\tau-\tau').
\end{split}
\ee
The factor $1/\tau$ stems from the Jacobian transformation of the space-time coordinates in Eq.~\eqref{eq:noisevar}. In order to find the asymptotic solutions we have to specify a functional form for the gauge field $A_\mu$. Since we are interested in the propagation of fluctuations that explicitly break the Bjorken symmetry, it is enough to consider an electric field that is a constant vector, i.e. the temporal component of the gauge field is a linear function that grows linearly with the distance $A^\mu=(\mc E\cdot\bx,0,0,0)$. 

The equations of motion for the Bjorken flow have a similar mathematical structure as their static counterparts~\eqref{eq:evoleqsCAB} so one can also obtain time-dependent asymptotic solutions. The diagonal components of the symmetric matrix $\C_{AB}$ admit the following asymptotic solution
\be
\label{eq:asymBjosol}
\C_{AA}=\frac{\C_0}{\tau}\,\chi_{AA}\left(1+\frac{2(1+v_a^2)+\tau\,\mc P_{AA}}{2\,D_{AA}\,{\bm K}^2\,\tau}
\right)\,,
\ee
where 
\be
\label{eq:PDelem}
\begin{split}
&\mc P_{\pm\pm}= -\frac{(2+v_a^2+\cos^2\theta_{\bbk})}{\tau}\,\pm\,\frac{\hbbk\cdot\mc E}{\bw}\frac{(1-v_a^2)}{v_a}\,,\\
&\mc P_{\bm{T}_1\bm{T}_1}=-\frac{2}{\tau}\,,\hspace{.5cm}\mc P_{\bm{T}_2\bm{T}_2}=-\frac{2\left(1+\sin^2\theta_{\bm K}\right)}{\tau}\,,\\
&\mc P_{dd}=-\frac{2}{\tau}\,,\\
&\mc D_{\pm\pm}= \frac{2}{3}\geta\,,\hspace{.5cm}
\mc D_{\bm{T}_l\bm{T}_l}=\geta\,,\hspace{.5cm}\mc D_{dd}=\,D\,.
\end{split}
\ee
We proceed as in Sect.~\ref{sec:sol} and obtain the asymptotic solution of the non-diagonal component $\C_{d\bm{T}_l}$
\be
\label{eq:asymcdTBjosol}
\C_{d\bm{T}_l}=-\frac{\chi_{d\bm{T}_l}}{\bw}\,\frac{\left(\hat{e}_{\bm{T}_l}\cdot\,\mc E\right)}{(D+\geta)\,{\bm K}^2\,}\,\frac{\C_0}{\tau}\,.
\ee
In the remainder of this section we outline the calculation of long time tails of the response functions in a Bjorken background. Some of the details are relegated to App.~\ref{sec:Greenfn}. In the static case we observed that the energy-momentum tails~\eqref{eq:corrtmn} are not affected by the chemical potential in the case of a conformal equation of state~\eqref{eq:confeos}. This is also the case for a Bjorken expansion, and we shall not repeat the calculation of Akamatsu et al.~\cite{Akamatsu:2016llw}. Instead, we will focus on stochastic corrections to the particle current. 

For the Bjorken flow the particle current in stochastic fluid dynamics is
\bs
\label{eq:fullcurrBjo}
\beal
\langle\,J^{\tau}\,\rangle&=\bd\,+\,\frac{\langle\,\left(\frg^x\right)^2+\,\left(\frg^y\right)^2+\left(\tau\frg^\varsigma\right)^2\rangle}{2\,\bw\,\Hb},\\
\langle\,J^{x}\,\rangle&=\sigma_0\mc E^x\,+\,\frac{\langle\,\delta n\,\frg^x\,\rangle}{\Hb}\,,\\
\langle\,J^{y}\,\rangle&=\sigma_0\mc E^y\,+\,\frac{\langle\,\delta n\,\frg^y\,\rangle}{\Hb}\,,\\
\langle\,\tau\,J^{\varsigma}\,\rangle&=\sigma_0\mc E^\varsigma\,+\,\frac{\langle\,\delta n\,(\tau\,\frg^\varsigma)\,\rangle}{\Hb}\,.
\end{align}
\es
where we have kept terms up to second order in fluctuating variables, and we have used that $\delta u^\tau=(\delta u^i)^2/2$. The correlation function  $\langle\phi_a\phi_b\rangle$ can be written in terms of the eigenstates of the acoustic matrix as
\be
 \langle \phi_a(x)\phi_b(x)\,\rangle= \tau\,\int \frac{d^3{\bm K}}{(2\pi)^3}(e_A)_a\,C_{AB}(\tau,{\bm K})\,(e_B)_b\,.
 \ee 
Using the asymptotic solutions for the correlation function~\eqref{eq:asymBjosol}-~\eqref{eq:asymcdTBjosol} we obtain the currents
\bs
\label{eq:fullcurrBjo-2}
\beal
\label{eq:Jt}
\left\langle\,J^\tau\,\right\rangle&=\bd\,+\,\frac{3\,T}{4\,\pi^2\bw}\int_0^\Lambda\,dK K^2\,+\,\frac{1}{2\bw}\Delta T^{\tau\tau}  \,,\\
\frac{\left\langle\,J^x\,\right\rangle}{\mc E^x}&=\frac{D_0}{\alpha_2}+\frac{1}{8\pi^2}\frac{T}{\bw^2}\frac{1}{\geta}\int_0^\Lambda\,dK\nonumber\\
& +\frac{1}{3\pi^2}\frac{T^2\,c_p}{\bw^2\Hb}\frac{1}{(D_0+\geta)}\int_0^\Lambda dK\,+\frac{\Delta J^x}{\mc E^x}\,,\\
\frac{\left\langle\,J^y\,\right\rangle}{\mc E^y}&=\frac{\left\langle\,J^x\,\right\rangle}{\mc E^x}\,,\\
\frac{\left\langle\,\tau J^\varsigma\,\right\rangle}{E^\varsigma}&=
\frac{D_0}{\alpha_2}+\frac{1}{8\pi^2}\frac{T}{\bw^2}\frac{1}{\geta}\int_0^\Lambda\,dK\nonumber\\
& +\frac{1}{3\pi^2}\frac{T^2\,c_p}{\bw^2\Hb}\frac{1}{(D_0+\geta)}\int_0^\Lambda dK+\frac{\tau\Delta J^\varsigma}{E^\varsigma}\,,
\end{align}
\es
In Eq.~\eqref{eq:Jt} we recognized that the quadratic perturbations of the momentum flux are uniquely related with $\delta T^{\tau\tau}$ (see Eq.~(67a) in Ref.~\cite{Akamatsu:2016llw}). The terms $\Delta J^\mu$ are the finite residual contributions to the particle current which are cutoff independent. The ultraviolet divergences are absorbed into the renormalized macroscopic hydrodynamic variables and parameters, i.e.,
\bs
\label{eq:renorvarBjo}
\beal
n&=\bd(\Lambda)\,+\,\frac{1}{4\pi^2}\frac{T}{\bw}\Lambda^3\,,\\
D&=D_0+\alpha_2\frac{T}{\bw^2}\frac{\Lambda}{\pi^2}\left(\frac{1}{8\geta}\,+\,\frac{T\,c_p}{3\Hb(D_0+\geta)}\right)\,,
\end{align}
\es
where the previous expressions are determined by comparing orders with the same power law $\tau$ dependence. The renormalized diffusion coefficient matches exactly the static homogeneous one~\eqref{eq:renor-D}. This also holds for the renormalized sound attenuation lenght $\geta$~\cite{Akamatsu:2016llw}. The second terms in the RHS of Eqs.~\eqref{eq:renorvarBjo} correspond to the bare and cutoff dependent terms. The cutoff dependent pieces, $\mc O(\Lambda)$ and $\mc O(\Lambda^3)$ respectively, are the contributions of the hydrodynamic fluctuating fields below the cutoff $\Lambda$. The cutoff dependent terms can be obtained directly from the asymptotic solutions of the symmetric matrix $\mc C_{AB}$, Eqs.~\eqref{eq:asymBjosol}-\eqref{eq:asymcdTBjosol}. The bare quantities captures the macroscopic thermodynamic equilibrium  properties of the fastest modes which have been already integrated out above the cutoff scale $\Lambda$. 

\subsection{Finite residual contributions of the particle current}
\label{subsec:finite}

In this section we determine finite cutoff-independent contributions to the particle current. We follow closely the procedure outlined in~\cite{Akamatsu:2016llw}. We compute the general solution for the correlation function $C_{AB}$ and focus on those terms that are independent of the initial conditions. We remove the leading divergence identified in the previous section, and compute the remaining finite terms. The technical details of this calculation are  presented in appendix \ref{sec:Greenfn}. 

From the formalism discussed in Sect.~\ref{sec:sol} the general solution of the evolution equation for the correlation matrix $\C_{AB}$ is
\be
\label{eq:CABgensol}
\begin{split}
\mc C_{AB}&= \mc U(\tau,\tau_0,{\bm K})\C_{AB}\left(\tau_0,{\bm K}\right)\,\\
&+\,\int_{\tau_0}^\tau\,d\tau'\,\mc U(\tau',\tau_0,{\bm K})\,g(\tau,{\bm K})\,,
\end{split}
\ee
with $\mc U(\tau,\tau',{\bm K})=e^{-\int_{\tau'}^\tau\,ds\,f(s,{\bm K})}$ where the term  $f(\tau,{\bm K})$ corresponds to the linear terms of the correlation function in Eqs.~\eqref{eq:EOMBcorrjo}. $f(\tau,{\bm K})$ takes into account the dissipative and external forcing sources due to either the expansion rate or the constant electric field. The term $g(\tau,{\bm K})$ is either the equilibrium value of the correlation function (for the diagonal correlators $\C_{AA}$) or the couplings between the external electric field and correlators $\C_{dd}$ and $C_{{\bm T}_l{\bm T}_l}$ (for the non-diagonal correlators $C_{d{\bm T}_l}$). As we pointed out previously, the first line of the solution~\eqref{eq:CABgensol} describes the decay of the initial condition for the correlator while the second line is the genuine contribution coming from hydrodynamic fluctuations~\cite{zwanzignonequilibrium}. We shall focus in the rest of this section only on the hydrodynamic fluctuating contributions.

Using the exact solution~\eqref{eq:CABgensol} and subtracting the divergent piece, we obtain the finite contribution to the temporal and spatial components of the particle current
\begin{widetext}
\bs
\label{eq:finJ}
\beal
\label{eq:deltaJt}
\Delta\,J^\tau&= \frac{1}{2\bw}\Delta T^{\tau\tau}=\frac{T}{\bw}\,\frac{0.04808}{\left(\geta\,\tau\right)^{3/2}}\,\,,\\
\frac{\Delta\,J^x}{\mc E^x}=\frac{\Delta\,J^y}{\mc E^y}&=-\frac{T\,\tau}{\bw^2}\left[\frac{0.0266}{(\gamma_\eta\tau)^{3/2}}\,+\,\frac{T\,c_p}{\Hb}\frac{0.03558}{[(D+\gamma_\eta)\tau]^{3/2}}\right]\,,\\
\frac{\Delta\,J^\varsigma}{\mc E^\varsigma}&=-\frac{T\,\tau}{\bw^2}\left[\frac{0.04282}{(\gamma_\eta\tau)^{3/2}}\,+\,\frac{T\,c_p}{\Hb}\frac{0.008}{[(D+\gamma_\eta)\tau]^{3/2}}\right]\,.
\end{align}
\es  
\end{widetext}
The coefficients appearing in the RHS of the previous expressions were calculated numerically as indicated in App.~\ref{sec:Greenfn}. 

\section{Conclusions}
%
%
%
\onecolumngrid
\begin{widetext}
\begin{table}[h]
\begin{center}
\renewcommand{\arraystretch}{1.3}
\begin{tabular}{ | c | c | c | c | }
  \cline{2-4} 
 \multicolumn{1}{ c| }{}  & {\vphantom{\LARGE Ap} Diagrammatic} &  {\vphantom{\LARGE Ap} Hydrokinetics} &  {\vphantom{\LARGE Ap} Hydrokinetics}
 \\ [1mm]  \multicolumn{1}{ c| }{}  & {\vphantom{\LARGE Ap} (Static)} &  {\vphantom{\LARGE Ap} (Static)} &  {\vphantom{\LARGE Ap} (Bjorken)}
 \\ [1mm] \hline \hline 
 {\vphantom{\LARGE Ap} $p(\Lambda)$} & $\pro-\frac{1}{15\pi^2}\,T\,\Lambda^3$ &  $\pro+\frac{1}{6\pi^2}\,T\,\Lambda^3$ &  $\pro+\frac{1}{6\pi^2}\,T\,\Lambda^3$\\ [2mm] \hline 
 {\vphantom{\LARGE Ap} $n(\Lambda)$} & 
 $\bd-\frac{1}{10\pi^2}\,\frac{T}{\bw}\,\Lambda^3$ &  $\bd+\frac{1}{4\pi^2}\,\frac{T}{\bw}\,\Lambda^3$ &  $\bd+\frac{1}{4\pi^2}\,\frac{T}{w}\,\Lambda^3$
 \\ [2mm] \hline 
 $\gamma_\eta(\lambda)$ & $\geta+\frac{\Lambda}{\geta}\frac{17}{120\pi^2}\frac{T}{\Hb}$
  & $\geta+\frac{\Lambda}{\geta}\frac{17}{120\pi^2}\frac{T}{\Hb}$ &$\geta+\frac{\Lambda}{\geta}\frac{17}{120\pi^2}\frac{T}{\Hb}$ \\ [3mm] \hline
 {\vphantom{\LARGE Ap}$D(\Lambda)$} &  $D_0+\,\alpha_2\,\frac{T}{\bw^2}\,\frac{\Lambda}{\pi^2}\left(\frac{1}{8\,\geta}+\frac{T\,\,c_p}{3\,\Hb\,(D+\geta)}\right)$ 
  & 
 $D_0+\,\alpha_2\,\frac{T}{\bw^2}\,\frac{\Lambda}{\pi^2}\left(\frac{1}{8\,\geta}+\frac{T\,\,c_p}{3\,\Hb\,(D+\geta)}\right)$ & 
 $D_0+\,\alpha_2\,\frac{T}{\bw^2}\,\frac{\Lambda}{\pi^2}\left(\frac{1}{8\,\geta}+\frac{T\,\,c_p}{3\,\Hb\,(D+\geta)}\right)$ 
  \\  [2mm] \hline 
\end{tabular}
\caption{\label{tab1}  Renormalized thermodynamic variables and transport coefficients due to hydrodynamic fluctuations in different approaches.  } 
\end{center}
\end{table}
\end{widetext}
\twocolumngrid
In this work we have studied the role of hydrodynamic fluctuations in a conformal fluid with a conserved $U(1)$ charge. The hydrokinetic formalism~\cite{Akamatsu:2016llw,Akamatsu:2017rdu,Akamatsu:2018vjr} has been extended to the regime of finite chemical potential and/or baryon density. The out-of-equilibrium hydrodynamic fluctuating contributions to the particle current renormalize the particle density and the diffusion coefficient. We have verified that the one-loop correction to the Green functions derived from the hydrokinetic and diagrammatic formalisms match exactly in the case of a static homogeneous background. The long time tails of the particle current are generated by contributions associated with sound modes and a mixture of shear and diffusive modes. The sound mode contribution is proportional to $\gamma_\eta^{-3/2}$ while the shear-diffusive one is proportional to $(D+\gamma_\eta)^{-3/2}$.  

Hydrodynamic fluctuations imply the existence of lower bounds on the transport coefficients. These bounds are purely classical and they are not universal. In fact, the lower bounds depend on the thermodynamic properties and the typical momentum scale at which hydrodynamics breaks down. The lower bound on the diffusion coefficient is enhanced when the value of the ratio of shear viscosity to entropy density is small and thus, the experimental determination of this type of transport coefficient has the potential to determine the impact of hydrodynamic fluctuations in relativistic heavy ion collisions.

In the case of a background undergoing boost invariant Bjorken flow the constitutive relations for the particle current in the presence of hydrodynamic fluctuations are
\begin{widetext}
\bs
\beal
\langle\,J^\tau\,\rangle&= n(\Lambda)\,+\,\frac{T}{\bw}\,\frac{0.04808}{\left(\geta\,\tau\right)^{3/2}}\,\,,\\
\frac{\langle\,J^x\,\rangle}{\mc E^x}=\frac{\langle\,J^y\,\rangle}{\mc E^y}&=\frac{D(\Lambda)}{\alpha_2}-\,\frac{T\,\tau}{\bw^2}\left[\frac{0.0266}{(\gamma_\eta\tau)^{3/2}}\,+\,\frac{T\,c_p}{\Hb}\frac{0.03558}{[(D+\gamma_\eta)\tau]^{3/2}}\right]\,,\\
\frac{\langle\,J^\varsigma\,\rangle}{\mc E^\varsigma}&=\frac{D(\Lambda)}{\alpha_2}-\,\frac{T\,\tau}{\bw^2}\left[\frac{0.04282}{(\gamma_\eta\tau)^{3/2}}\,+\,\frac{T\,c_p}{\Hb}\frac{0.008}{[(D+\gamma_\eta)\tau]^{3/2}}\right]\,.
\end{align}
\es 
\end{widetext}
where the particle density as well as the shear and diffusive transport coefficients are renormalized quantities. In Table~\ref{tab1} we summarize the properties of hydrodynamic fluctuations obtained from the different approaches in the static and expanding cases. The renormalized transport coefficients obtained in the diagrammatic approach coincide with the hydrokinetic approach in both homogeneous and expanding fluids. There is a mismatch between the hydrokinetic and diagrammatric results for the cubic divergences that enter the pressure and particle density \cite{Chafin:2012eq,Kovtun:2011np}. This mismatch arises from the approximations made in the propagators of the sound modes, which affect the UV behavior of the loop integrals. 

We are now in a position to estimate the relative size of fluctuation corrections to the $U(1)$ current. For simplicity we will focus on the spatial components $\langle J^x\rangle/{\cal E}^x=\langle J^y\rangle/{\cal E}^y$. As an estimate of the renormalized conductivity $\sigma=D/\alpha_2$ we will use the lower bound~\eqref{eq:lowD} derived in Sect.~\ref{sec:lower}. This has the virtue that the gradient term and the fluctuation correction scale in the same way with the enthalpy per particle $\bar{w}$. We use $\lambda=1/4$ as suggested by simple kinetic estimates. Both the bound on the gradient term and the fluctuation corrections are enhanced by a term proportional to $c_p$. For simplicity we ignore these terms. We express $\eta/s$ in units of $1/(4\pi)$, and the dimensionless proper time $\tau T$ in units of a typical freezeout time $\tau T\simeq 4.5$. We find 
\be
\frac{\langle J^x\rangle}{{\cal E}^x}= \frac{T^3(s/\eta)^2}{4\pi^2\bar{w}^2}
  \left\{
  1 - 0.14 \left(\frac{4\pi}{s/\eta}\right)
           \left(\frac{4.5}{\tau T}\right)^{1/2} \right\}\, , 
\ee
indicating that fluctuations are important, but do not overwhelm the leading term in the gradient expansion at freezeout.

There are a number of issues that can be studied in the future. On the technical side it would be interesting to study convergence properties and higher order asymptotics of the long time expansion in the presence of fluctuations. This is related to resurgent asymptotics, which has received a considerable amount of attention in the context of both kinetic theory and holographic dualities. From a phenomenological point of view it is crucial to extend this formalism to include critical behavior~\cite{Stephanov:1998dy,Akamatsu:2018vjr,Bluhm:2016byc,Singh:2018dpk,Hirano:2018diu}. Finally it would be interesting to study fluctuations in time-dependent backgrounds in non-relativistic fluids with possible applications to ultracold atomic gases or graphene~\cite{Chafin:2012eq,Romatschke:2012sf,Martinez:2017jjf,Kovtun:2014nsa,Chen-Lin:2018kfl}.

\acknowledgments
We would like to thank D.~Teaney, M.~Stephanov, J.~Kapusta, A.~Cherman, G.~Basar and Ho-Ung Yee for enlightening discussions, and the members of the Nuclear Theory Group at NCSU for advice on the regularization of divergent integrals. This work was supported in part by the US Department of Energy grant DE-FG02-03ER41260 and by the BEST (Beam Energy Scan Theory) DOE Topical Collaboration. 

\appendix

\section{Thermodynamic relations at finite chemical potential}
\label{app:thermo}

In this appendix we describe the thermodynamic relations of a relativistic fluids with a conserved $U(1)$ charge in the grand canonical potential $\Omega$ (see also Ref.~\cite{Floerchinger:2015efa}). Let us first introduce the Jacobian
\be
\label{eq:jacobian}
\frac{\partial\left(u,v\right)}{\partial\left(x,y\right)}=\begin{vmatrix}
\frac{\partial u}{\partial x}&\frac{\partial u}{\partial y}\\
\frac{\partial v}{\partial x}&\frac{\partial uv}{\partial y}
\end{vmatrix}
\ee
In the grand canonical potential the pressure is given by $p=-\Omega/V$. The entropy density and particle density are determined from the pressure through the following thermodynamical expressions
\be
s=\frac{\partial p}{\partial T}\biggl|_{\mu}\,,\hspace{2cm}
n=\frac{\partial p}{\partial \mu}\biggl|_{T}\,.
\ee
The energy density is obtained from the Gibbs-Duhem relation 
\be
\begin{split}
\epsilon&=\mu\,n\,+\,T\,s\,-\,p\,,\\
&=\mu\frac{\partial p}{\partial \mu}\biggl|_{T}\,+\,T\,\frac{\partial p}{\partial T}\biggl|_{\mu}\,-\,p\,.
\end{split}
\ee
Let us then calculate the specific heat at constant volume  
\be
\label{eq:cv}
\begin{split}
c_V  = \frac{T}{V} \left(\frac{\partial S}{\partial T} \right)_{V,N} &= T \left( \frac{\partial s}{\partial T} \right)_n\,=\,T\,\left(\frac{\partial(s,n)}{\partial(T,\mu)}\right)\biggl/\,\frac{\partial n}{\partial \mu}\biggl|_T\,.
\end{split}
\ee
The isothermal and adiabatic compressibilities are
\bs
\label{eq:compr}
\beal
\kappa_T = & - \frac{1}{V} \left( \frac{\partial V}{\partial p} \right)_{T,N} = \frac{1}{n} \left( \frac{\partial n}{\partial p} \right)_T= \frac{1}{n^2} \frac{\partial n}{\partial\mu}\,,\\
\kappa_S = & - \frac{1}{V} \left(\frac{\partial V}{\partial p}\right)_{S,N} = \frac{1}{n} \left( \frac{\partial n}{\partial p} \right)_{s/n}\,
= \frac{ \frac{\partial(s,n)}{\partial (T,\mu)}}{s\frac{\partial(p,n)}{\partial(T,\mu)}-n\frac{\partial(p,s)}{\partial(T,\mu)}}\,.
\end{align}
\es
The thermal expansion coefficient is
\be
\label{eq:expcoeff}
\alpha = \frac{1}{V} \left( \frac{\partial V}{\partial T} \right)_{P,N} = - \frac{1}{n} \left( \frac{\partial n}{\partial T} \right)_P = \frac{1}{n^2} \left( \frac{\partial (p,n)}{\partial(T,\mu)}\right)\,,
\ee
Now, we can compute the specific heat at constant pressure $c_P$ by using the thermodynamic relation $c_P-c_V=T\alpha^2/\kappa_T$. Using Eqs.~\eqref{eq:cv}-\eqref{eq:expcoeff} we get as a result
\be
\label{eq:cp}
c_P=T\,\left(\frac{\partial(s,n)}{\partial(T,\mu)}\,+\,\frac{1}{n^2}\left(\frac{\partial (p,n)}{\partial (T,\mu)}\right)^2\right)\biggl/\,\frac{\partial n}{\partial \mu}\biggl|_T\,.
\ee
Notice that the thermodynamical relations
\be
\label{eq:thermrelat}
\begin{split}
\frac{c_P}{c_V} = \frac{\kappa_T}{\kappa_S}, \quad\quad & \kappa_T-\kappa_S = \frac{T \alpha^2}{c_P}\,, 
\end{split}
\ee
are satisfied. One can write the adiabatic speed of sound $v_a^2$ in terms of the coefficients calculated above. As a result one gets
\be
\label{eq:adiabva}
\begin{split}
v_a^2=\frac{\partial p}{\partial\epsilon}\biggl|_{s/n}\,&=\frac{1}{\kappa_S\,(\epsilon+p)},\\
&=\frac{c_P}{c_V}\,\frac{1}{\kappa_T\,(\epsilon+p)}\,=\,\frac{1}{\kappa_T\,(\epsilon+p)}\left(1+\frac{T\alpha^2}{c_V\,\kappa_T}\right)\,.
\end{split}
\ee
The isothermal speed of sound is
\be
c_s^2=\frac{\partial p}{\partial \epsilon}\biggl|_{T}=\frac{n}{T}\,\left(\frac{\partial n}{\partial T}\biggl|_{\mu/T}\right)^{-1}\,.
\ee
The coefficient $\alpha_1$~\eqref{eq:alpha1} can be rewritten as
\be
\label{eq:alpha1tmu}
\begin{split}
\alpha_1&=\left(\frac{\partial\,\mu}{\partial\,\epsilon}\right)_n-\frac{\mu }{T}\,\left(\frac{\partial\,T}{\partial\,\epsilon}\right)_n\,,\\
&=-\left(\frac{\partial n}{\partial T}\biggl|_{\mu/T}\right)\biggl/\left(T\,\frac{\partial\left(s,n\right)}{\partial\left(T,\mu\right)}\right)\,,\end{split}
\ee
Now, the coefficient $\alpha_2$~\eqref{eq:alpha2} is
\be
\label{eq:alpha2tmu}
\begin{split}
\alpha_2&=\left(\frac{\partial\,\mu}{\partial\,n}\right)_\epsilon-\frac{\mu }{T}\,\left(\frac{\partial\,T}{\partial\,n}\right)_\epsilon\,,\\
&=\,\left(T\frac{\partial s}{\partial T}\biggl|_{\mu/T}\,+\,\mu\frac{\partial n}{\partial T}\biggl|_{\mu/T}\,\right)\biggl/\left(T\,\frac{\partial\left(s,n\right)}{\partial\left(T,\mu\right)}\right)\,,
\end{split}
\ee
During the text we find the combination $\alpha_1\,+\,\alpha_2/w$ which can be written as
\be
\label{eq:comb}
\alpha_1\,+\,\frac{\alpha_2}{w}\,=\,\left(n\frac{\partial s}{\partial T}\biggl|_{\mu/T}-s\frac{\partial s}{\partial\mu}\biggl|_{\mu/T}\right)\biggl/\left(n\,\frac{\partial (s,n)}{\partial (T,\mu)}\right)
\ee
The coefficient $\beta_1$ is given by
\be
\label{eq:beta1}
\begin{split}
\beta_1=\frac{\partial p}{\partial\epsilon}\biggl|_{n}&=\left(\frac{\partial(p,n)}{\partial(T,\mu)}\right)\biggl/\left(T\,\frac{\partial (s,n)}{\partial (T,\mu)}\right)\,,\\
&=\frac{\alpha}{\kappa_T\,c_V}
\end{split}
\ee
We can also obtain the coefficient $\beta_2$ 
\be
\label{eq:beta2}
\begin{split}
\beta_2\,=\,\frac{\partial p}{\partial n}\biggl|_{\epsilon}&=\frac{1}{n\,\kappa_T\,c_V}\,\left(c_P-\alpha\,(\epsilon+p)\,\right)\,.
\end{split}
\ee
The susceptibility matrix $\chi_{ab}$ relates the variations between the hydrodynamical fields $\varphi_a=(\delta \epsilon,g_i,\delta n)$ (i=1,2,3) and the variations of the temperature $\delta T$, fluid velocity $\delta u_i$ and chemical potential $\delta\mu$ throught the relation $\varphi_a=\chi_{ab}\,\rho_b$ with $\rho_a=(\delta T/T,\delta u_i,\delta\mu-(\mu/T)\,\delta T)$~\cite{Kovtun:2012rj} and $\chi_{ab}$ given by
\be
\label{eq:chimat}
\chi=\begin{pmatrix}
T\left(\frac{\partial\epsilon}{\partial T}\right)_{\mu/T}&0&\left(\frac{\partial\epsilon}{\partial \mu}\right)_{T}\\
0&H\delta_{ij}&0\\
T\left(\frac{\partial n}{\partial T}\right)_{\mu/T}&0&\left(\frac{\partial n}{\partial \mu}\right)_{T}
\end{pmatrix}\,.
\ee
It follows that $\text{det}\,\chi=H^3\,\left(\chi_{11}\chi_{33}-\chi_{13}\chi_{31}\right)$. The inverse susceptibility matrix $\left(\chi^{-1}\right)_{ab}$ which is
\be
\label{eq:invchimat}
\chi^{-1}=\begin{pmatrix}
\frac{\chi_{55}}{\chi_{11}\chi_{55}-\chi_{15}\chi_{51}}&0&-\frac{\chi_{15}}{\chi_{11}\chi_{55}-\chi_{15}\chi_{51}}\\
0&\frac{1}{H}\delta_{ij}&0\\
-\frac{\chi_{51}}{\chi_{11}\chi_{55}-\chi_{15}\chi_{51}}&0&\frac{\chi_{11}}{\chi_{11}\chi_{55}-\chi_{15}\chi_{51}}
\end{pmatrix}
\ee
Onsager relations~\cite{onsager1,onsager2} require the susceptibility matrix $\chi_{ab}$~\eqref{eq:chimat} to be symmetric. This condition is satisfied in Eq.~\eqref{eq:chimat} because in the gran canonical potential $T\left(\partial n/\partial T\right)_{\mu/T}= \left(\partial \epsilon/\partial \mu\right)_{T}$ so $\chi_{15}=\chi_{51}$.  One can also show that $\chi_{11},\chi_{55}\geq 0$ as well as $\text{det}\,\chi\geq 0$ (see discussion in Sect. 2.5 of Ref.~\cite{Kovtun:2012rj}). Furthermore, for time-reversal invariant theories the Maxwell relations together with the Gibbs-Duhem relation lead to the following identities~\cite{Kovtun:2012rj} 
\bs
\label{eq:susident}
\beal
\beta_1\,\chi_{11}\,+\,\beta_2\,\chi_{51}&=\,H\,,\\
\beta_1\,\chi_{15}\,+\,\beta_2\chi_{55}&=\,n\,,\\
\alpha_1\,\chi_{11}\,+\,\alpha_2\,\chi_{51}&=\,0\,,\\
\alpha_1\,\chi_{15}\,+\,\alpha_2\,\chi_{55}&=\,1\,.
\end{align}
\es
Thus one can express the thermodynamic coefficients $\alpha_{1,2}$, Eqs.~\eqref{eq:alpha1tmu}-\eqref{eq:alpha2}, and $\beta_{1,2}$, Eqs.~\eqref{eq:beta1}-~\eqref{eq:beta2},  in terms of the coefficients of the susceptibility matrix $\chi_{ab}$ and viceversa.  

Now for the conformal EOS given by
\be
\label{eq:idealeos}
P(T,\mu)=T^4\,g(\mu/T),. 
\ee
the energy, particle density and entropy densities are given by respectively
\bs
\label{eq:thermquant}
\beal
\epsilon&=3\,T^4\,g(\mu/T) \,,\\
n&=T^3\,g'(\mu/T)\,,\\
s&=T^2\left(4Tg(\mu/T)-\mu g'(\mu/T)\right)\,,
\end{align}
\es
where $g'(\mu/T)$ is the derivative respect to the argument. It follows that $\epsilon=3 p$ so $v_a^2=1/3$ and thus, the enthalpy density $H=\epsilon+p=4p$. Using the results of the previous sections we calculate the thermodynamical coefficients $\beta_1$, $\beta_2$, $\alpha_1$ and $\alpha_2$ for the EOS~\eqref{eq:idealeos} which gives us respectively
\bs
\beal
\beta_1&=\frac{1}{3}\,,\\
\beta_2&=0\,,\\
\label{eq:a1conf}
\alpha_1&=\,\frac{\,g'(\mu/T)}{T^3\left(3\left(g'(\mu/T)\right)^2-4\,g(\mu/T)\,g''(\mu/T)\right)},\\
\label{eq:a2conf}
\alpha_2&=\,\frac{4\,g(\mu/T)}{T^2\left(4\,g(\mu/T)\,g''(\mu/T)-3\left(g'(\mu/T)\right)^2\right)}
\,.
\end{align}
\es
It is straighforward to show that for the EOS~\eqref{eq:idealeos} the combination $\alpha_1+\alpha_2/w=0$.

\section{Long time tails of the $\langle\,J_i\,J_k\,\rangle$ correlator}
\label{app:supp}

In this section we present the calculation of the time symmetrized (unordered) correlator $\langle\,J\,J\,\rangle$ correlator in the approach used by Kovtun and Yaffe~\cite{Kovtun:2003vj}. In this method the correlators for the particle, energy and momentum densities are obtained from the solutions of linearized hydrodynamics. By considering the same approach we show in this appendix how to obtain the long time tails of the $\langle JJ\rangle$ correlator when one uses $(\dep/v_a,g_i,\dq)$ as the set of hydrodynamical fluctuating variables. For the EOS~\eqref{eq:confeos} the calculation of the $\langle\left\{T_{ij},T_{kl}\right\}\rangle$ correlator is the exactly the same so we refer to the interested reader to the original reference~\cite{Kovtun:2003vj}. 

Now, in the absence of external and noise sources the linearized hydrodynamical equations for $(\dep/v_a,\gL,\bm{\frg}_{T_l},\dq)$ (with $l=1,2$)~\eqref{eq:NSLeqs} are
\bs
\label{eq:KYeqs}
\beal
\partial_t\left(\frac{\dep}{v_a}\right)\,+\,ik\,v_a\,\gL&=0\,,\\
\partial_t\gL\,+\,ik\,v_a\,\frac{\dep}{v_a}\,+\,\frac{4}{3}\,\geta\,k^2\,\gL&=0\,,\\
\partial_t\bm{\frg}_{T_l}\,+\,\,\geta\,k^2\,\bm{\frg}_{T_l}&=0\,,\hspace{0.5cm}\text{with l=1,2}\,,\\
\partial_t\dq\,+\,D\,k^2\,\dq&=0\,.
\end{align}
\es
For $t\geq 0$ and in the long wavelength limit ($\geta\,k\ll v_a$) the solutions for the fluctuating fields $\Psi_a=\left(\dep/v_a,\gL,\bm{\frg}_{T_l},\dq\right)$ are 
\be
\label{eq:Greentime}
\Psi_a(t,\bk)=\mc G_{ab}(t,\bk)\Psi_b(0,\bk)\,,
\ee
where $\Psi_a(0,\bk)$ are the  set of initial conditions for each hydrodynamic fluctuating field. In the previous expression we recognize $G_{\phi_a\phi_b}(t,\bk)$ as the two point symmetric correlation matrix defined as~\cite{Kovtun:2003vj}  
\begin{equation*}
\mc G_{\phi_a\phi_b}(t)=\int_{\bm x}\,e^{-i\bk\cdot\bx}\,\biggl\langle\,\frac{1}{2}\left\{\phi_a(t,\bx),\phi_b(0,\bm{0})
\right\}\biggr\rangle\,.
\end{equation*}
The explicit functional form of each of the components of $G_{\phi_a\phi_b}(t,\bk)$ appearing in Eq.~\eqref{eq:Greentime} are
\bs
\label{eq:solsKYeqs}
\beal
\mc G_{\dep\dep}(t,\bk)=\mc G_{\gL\gL}(t,\bk)&= e^{-\frac{2}{3}\geta k^2 t}\cos\left(kv_a t\right)\,,\\
\mc G_{\dep\gL}(t,\bk)=\mc G_{\gL\dep}(t,\bk)&=-i\,e^{-\frac{2}{3}\geta k^2 t}\,\sin\left(kv_a t\right)\,,\\
\mc G_{\frg_{T_l}\,\frg_{T_l}}(t,\bk)&=e^{-\geta\,k^2 t}\,,\\
\mc G_{\dq\dq}(t,\bk)&=e^{-D\,k^2 t}\,,
\end{align}
\es
In order to make connection with correlators one follows the well known procedure of linear response theory~\cite{forster1995hydrodynamic}.
The real time symmetric $\langle\,J\,J\,\rangle$ correlator is determined by using these solutions as follows
\begin{widetext}
\be
\label{eq:JJcorr-2}
\begin{split}
\frac{1}{\mathcal{V}}\left\langle\,\frac{1}{2}\left\{\,J^i(t),J^j(0)\,\right\}\right\rangle\,&=\frac{1}{\Hb^2}\,\int\,d^3\bm{x}\,\left\langle\frac{1}{2}\left\{\dbd(t,\bm{x}),\dbd(0,\bm{0})\right\}\right\rangle\,\left\langle\frac{1}{2}\left\{\frg^i(t,\bm{x}),\frg^j(0,\bm{0})\right\}\right\rangle\,,\\
&=\frac{T^2}{\Hb\,\bw^2}\,\int\,\frac{d^3{\bk}}{(2\pi)^3}\,\left[\frac{\Hb}{v_a^2}\mc A_{\dep\dep}(t,\bk)\,+T\,c_p\,\mc A_{\dq\dq}(t,\bk)\right]\,\\
&\hspace{1cm}\times\,\left[\hbk_i\hbk_j\,\mc A_{\gL\gL}(t,-\bk)\,+\,\sum_{l=1}^2\,\left(\etr l\right)_i\left(\etr l\right)_j\,\mc A_{\frg_{\bm{T}_l}\frg_{\bm{T}_l}}(t,-\bk)\right]\,,\\
&\approx\,\delta_{ij}\,\frac{T^2}{\bw^2}\left[\frac{1}{12\,\pi^{3/2}}\frac{T\,c_p}{\Hb}\frac{1}{\left[(D+\geta)\,t\right]^{3/2}}\,+\,\frac{1}{16}\left(\frac{3}{4\pi}\right)^{3/2}\frac{1}{\left(\geta t\right)^{3/2}}\right]\,,
\end{split}
\ee
\end{widetext}
where $\mc V$ is the spatial volume and we used $v_a^2=1/3$. We explicitly used that the mean square fluctuations in momentum $\langle p^2\rangle/[3\,\mc V]=\Hb\,T$ and the mean square fluctuations of energy (and thus, pressure) $C=\Hb T/v_a^2$~\cite{Kovtun:2003vj}. Furthermore, the mean square fluctuation of the heat energy density is determined via thermodynamics, i.e. $\langle\,\left[\dq(0,\bm{0})\right]^2\,\rangle\,=\,T^2\, c_p$ being $c_p$ being the specific heat at constant pressure. Furthermore, In Eqs.~\eqref{eq:JJcorr-2} we performed the following integrals over the azymuthal angle are
\begin{subequations}
\beal
\int\,d\Omega\,\hbk_i\hbk_j&=\frac{\delta_{ij}}{3}\,\\
\int\,d\Omega \left(\etr 1\right)_i\left(\etr 1\right)_j&=
\begin{cases}
\frac{\delta_{ij}}{2}&\hspace{.5cm} \text{if}\,\,i,j=1,2\\
0&\hspace{.5cm} \text{if}\,\,i,j=3\,,
\end{cases}\\
\int\,d\Omega \left(\etr 2\right)_i\left(\etr 2\right)_j&=
\begin{cases}
\frac{1}{6}&\hspace{.5cm} \text{if}\,\,i,j=1,2\\
\frac{2}{3}&\hspace{.5cm} \text{if}\,\,i,j=3
\end{cases}
\end{align}
\end{subequations}
with $d\Omega\equiv d(\sin\theta)d\phi/(4\pi)$.

Thus the $\langle JJ\rangle$ long time tails arise from the mixed shear-heat mode and the sound-sound mode. This results differs from the conformal single charged fluid case studied in Sect. II of Ref.~\cite{Kovtun:2003vj} (see Eq. (37)). In their approach the authors explicitly neglect the couplings between the particle density fluctuations and other hydrodynamical fluctuating fields so the behaviour of the density-density correlator is purely diffusive. This does not hold in our case and other non-relativistic systems~\cite{Chafin:2012eq,Martinez:2017jjf,pomeau1975time}. 

On the other hand we can reconstruct the two point $\left\langle JJ\right\rangle$ correlator in frequency space by taking the one sided Fourier transform in Eq.~\eqref{eq:JJcorr-2}. Thus, we get
\begin{widetext}
\be
\label{eq:JJcorr}
\begin{split}
\left\langle\,\frac{1}{2}\left\{J^i(\omega,\bm{0}),J^j(0,\bm{0})\right\}\right\rangle&=\, \frac{1}{\mathcal{V}}\,\int_0^\infty dt\, e^{i\omega t}\left\langle\,\frac{1}{2}\left\{\,J^i(t),J^j(0)\,\right\}\right\rangle\,\,,\\
&=\delta_{ij}\left(\frac{T}{\bw}\right)^2\,\left[\frac{\Lambda}{\pi^2}\left(\frac{1}{8\geta}+\frac{T\,\,c_p}{3\,\Hb\,(D+\geta)}\right)\right.\,\\
&\left.\hspace{2cm}+\,\frac{(1-i)}{\sqrt{2}}\frac{\sqrt{\omega}}{\pi}\left(\frac{1}{12}\left(\frac{3}{4}\right)^{3/2}\frac{1}{\geta^{3/2}}\,+\,\frac{T\,c_p}{6\,\Hb}\frac{1}{(D+\geta)^{3/2}}\right)
\right]
\end{split}
\ee
\end{widetext}
where again one considers that close to the equilibrium the fluctuations of the pressure and heat energy density as well as the pairs of momentum flux components are independent~\cite{landauv9} and the distribution of fluctuations is Gaussian~\cite{Kovtun:2003vj}. 

\section{Green functions and residual contributions to $J^\mu$}
\label{sec:Greenfn}

In order to calculate the independent cutoff contributions to the charge current~\eqref{eq:fullcurrBjo-2} it is convenient to rewrite all the equations of motion in terms of the response ratio fuction $\mc R_{AB}$ defined as~\cite{Akamatsu:2016llw}
\be
\label{eq:Rfin}
\mc R_{AB}=\frac{\mc C_{AB}}{\mc C_0/\tau}\,,
\ee
which measures the deviation of the full response function from its equilibrium value. The equations for the different response ratio functions read as
\be
\label{eq:Reqn}
\begin{split}
\partial_\tau \mc R_{AA}\,+\,2\mc D_{AA}&{\bm K}^2\,\mc R_{AA}= 2\chi_{AA}\mc D_{AA}{\bm K}^2\\
&+\left[2(1+v_a^2)+\tau\mc P_{AA}\right]\frac{\mc R_{AA}}{\tau}\,,
\end{split}
\ee
with $A={\pm,{\bm T}_l}$, the terms  $\mc P_{AA}$ and $\mc D_{AA}$ are given by Eqs.~\eqref{eq:PDelem} and we use that $\partial_\tau\,\left[\mc C_0(\tau)/\tau\right]\approx\,-2\left(1+v_a^2\right)\,\left[\mc C_0(\tau)/\tau^2\right]$.
The general solutions for $\mc R_{AA}$ are
\begin{widetext}
\bs
\label{eq:RAAsol}
\begin{align}
\label{eq:Rpmsol}
\mc R_{\pm\pm}&(\tau,{\bm K})=
\mc G_{\pm\pm}(\tau,\tau_0,{\bm K})\,R_{\pm\pm}(\tau_0,{\bm K}_0)\,
+\,2\,\int_{\tau_0}^\tau\,d\tau'\,\mc G_{\pm\pm} (\tau',\tau,{\bm K})\,\left[\frac{4}{3}\gamma_\eta(\tau'){\bm K^2}\,\pm\,\frac{\hat{\bm K}\cdot \mc E}{\bw(\tau')}\left(\frac{1-v_a^2}{v_a}\right)
\right]
\,,\\
\mc R_{{\bm T}_l{\bm T}_l}&(\tau,{\bm K})= \mc G_{{\bm T}_l{\bm T}_l} (\tau,\tau_0,{\bm K})\,R_{{\bm T}_l{\bm T}_l}(\tau_0,{\bm K}_0)\,+\,2\,\int_{\tau_0}^\tau\,d\tau'\,\mc G_{{\bm T}_l{\bm T}_l} (\tau',\tau,{\bm K})\,\gamma_\eta(\tau'){\bm K^2}\,,\\
\mc R_{dd}&(\tau,{\bm K})=\mc G_{dd} (\tau,\tau_0,{\bm K})\,R_{dd}(\tau_0,{\bm K}_0)\,+\,2\,\int_{\tau_0}^\tau\,d\tau'\,\chi_{dd}\,\mc G_{dd} (\tau',\tau,{\bm K})\,D(\tau'){\bm K^2}\,,
\end{align}
\es
\end{widetext}
where ${\bm K}_0\equiv {\bm K}(\tau_0)$ and $\mc G_{AA} (\tau,\tau_0,{\bm K})$ is the Green function. An explicit calculation of the Green functions yields to the following expressions
\bs
\label{eq:Greenfuncs}
\beal
\label{eq:Greensound}
&G_{\pm\pm} (\tau',\tau,{\bm K})= \frac{1}{t^{v_a^2}}\frac{e^{- \frac{4}{3}\,{\bm r}_s^2 B(t,\theta_{\bm K})}}{\sqrt{A(t,\theta_{\bm K})}}\,
\,,\\
&G_{{\bm T}_1{\bm T}_1} (\tau',\tau,{\bm K})= \frac{1}{t^{2v_a^2}}\,e^{- 2\,{\bm r}_s^2 B(t,\theta_{\bm K})} \,,\\
&G_{{\bm T}_2{\bm T}_2} (\tau',\tau,{\bm K})= \frac{1}{t^{2v_a^2-2}}\, A(t,\theta_{\bm K})\,e^{- 2\,{\bm r}_s^2 B(t,\theta_{\bm K})} \,,\\
&G_{dd} (\tau',\tau,{\bm K})=\frac{1}{t^{2v_a^2}} e^{- 2\,{\bm r}_d^2 B(t,\theta_{\bm K})}\,,
\end{align}
\es
with $t=\tau'/\tau$, ${\bm r}_s=(\gamma_\eta(\tau)\tau)^{1/2}\,{\bm K}$ and ${\bm r}_d=(D(\tau)\tau)^{1/2}\,{\bm K}$. Moreover we define the following functions
\bs
\label{eq:A-Bfunc}
\beal
A(t,\theta_{\bm K})&=\sin^2\theta_{\bm K}+\frac{\cos^2\theta_{\bm K}}{t^2}\,,\\
\label{eq:Bfunc}
B(t,\theta_{\bm K})&= \frac{\sin^2\theta_{\bm K}}{1+v_a^2}\left(1-t^{1+v_a^2}\right)+\frac{\cos^2\theta_{\bm K}}{1-v_a^2}\left(\frac{1}{t^{1-v_a^2}}-1\right)\,,
\end{align}
\es
In the derivation of Eqs.~\eqref{eq:Greenfuncs} we used the Bjorken scaling of the temperature and finite chemical potential scale like $\tau^{-1/3}$ in the ideal fluid limit~\cite{Floerchinger:2015efa}. We approximate $\left(\hat{\bm K}\cdot \mc E\right)\,\mc R_{\pm\pm}\approx\,\left(\hat{\bm K}\cdot \mc E\right)$ in the LHS of Eq.~\eqref{eq:Reqn} when $A=\pm$. This is justified since we are interested only in the linear response of the . Thus, this term enters as an independent source in the LHS of the corresponding equation for $\mc R_{\pm\pm}$. 

By following a similar procedure we find that the differential equation for the response ratio functions $\mc R_{d{\bm T}_l}=\mc C_{d{\bm T}_l}/\left(\mc C_0/\tau\right)$ are
\be
\label{eq:RdTeq}
\begin{split}
\partial_\tau \mc R_{d{\bm T}_l}\,+\,(D_0+\geta)&{\bm K}^2\,\mc R_{d{\bm T}_1}=\frac{P_{d{\bm T}_l}}{\tau}\,\mc R_{d{\bm T}_l}\\
&-\frac{\hat{e_{{\bm T}_l}\cdot\mc E}}{\bw}\,\left(\,\mc R_{{\bm T}_l{\bm T}_l}+\mc R_{dd}\right)\,,
\end{split}
\ee
with $\mc P_{d{\bm T}_l}$ given by
\be
\mc P_{d{\bm T}_1}=2v_a^2\,,\hspace{1.5cm}
\mc P_{d{\bm T}_2}=2(v_a^2-\sin^2\theta_{\bm K})\,.
\ee
The solution of Eq.~\eqref{eq:RdTeq} is
\be
\label{eq:RdTsol}
\begin{split}
\mc R_{d{\bm T}_l}&(\tau,{\bm K})=\mc G_{d{\bm T}_l}(\tau,\tau_0,{\bm K})\mc R_{d{\bm T}_l}(\tau_0,{\bm K})\\
&-\int_{\tau_0}^\tau\,d\tau'\,\frac{\left(\hat{e}_{{\bm T}_l}\cdot\mc E\right)}{\bw(\tau')}\,\mc G_{d{\bm T}_l}(\tau',\tau_0,{\bm K})\,\left(\mc R_{{\bm T}_l{\bm T}_l}+\mc R_{dd}\right)\,.
\end{split}
\ee 
The explicit form of the Green functions $\mc G_{d{\bm T}_l}$ are
\bs
\label{eq:dTGreenfuncs}
\beal
\mc G_{d{\bm T}_1}&= \frac{1}{t^{2v_a^2}}\,e^{-\frac{{\bm K}^2}{K_{d{\bm T}}^2}\,B(t,\theta_{\bm K})}\,,\\
\mc G_{d{\bm T}_2}&= \frac{1}{t^{2v_a^2-2}}\,A(t,\theta_{\bm K})\,e^{-\frac{{\bm K}^2}{K_{d{\bm T}}^2}\,B(t,\theta_{\bm K})}\,.
\end{align}
\es
with $K_{d{\bm T}}=[(D+\gamma_\eta)\tau]^{-1/2}$. In order to derive the previous expressions we used explicitly that the Wiedemann-Franz law holds, i.e., $D\sim\gamma_\eta$~\cite{Jaiswal:2015mxa,Son:2006em}. 

Now we define the residual function $\mc R^{(r)}_{AB}$ as the difference between the exact solution~\eqref{eq:RAAsol}~\cite{Akamatsu:2016llw} and its leading asymptotic series term (which reads off directly from Eqs.~\eqref{eq:asymBjosol} or Eq.~\eqref{eq:asymcdTBjosol}), i.e., 
\be
\label{eq:residualR}
\mc R^{(r)}_{AB}= \mc R^{(ex.)}_{AB}- \mc R^{(asym.)}_{AB}\,.
\ee
In terms of these functions the finite residual contributions to the particle current components are given by the following expressions
\begin{widetext}
\bs
\label{eq:resJ}
\beal
\label{eq:resJx}
\frac{\Delta\,J^x}{\mc E^x}&=\frac{T}{\mc E^x\,\bw}\,\int\frac{dK\,d(\cos\theta_{\bm K})\,d\phi_{\bm K}\,K^2}{(2\pi)^3}\left[\frac{\sin\theta_{\bm K}\cos\phi_{\bm K}}{4v_a}\left(\mc R^{(r)}_{++}\,-\,\mc R^{(r)}_{--}\right)\,+\,\sin\phi_{\bm K}\mc R^{(r)}_{d{\bm T}_1}\,-\,\cos\theta_{\bm K}\cos\phi_{\bm K}\mc R^{(r)}_{d{\bm T}_2}\right]\,,\\
\label{eq:resJvar}
\frac{\tau \Delta\,J^\varsigma}{\mc E^\varsigma}&=\frac{T}{\mc E^\varsigma\,\bw}\,\int\frac{dK\,d(\cos\theta_{\bm K})\,d\phi_{\bm K}\,K^2}{(2\pi)^3}\left[\frac{\cos\theta_{\bm K}}{4v_a}\left(\mc R^{(r)}_{++}\,-\,\mc R^{(r)}_{--}\right)\,+\,\sin\theta_{\bm K}\,\mc R^{(r)}_{d{\bm T}_2}\right]\,.
\end{align}
\es  
\end{widetext}
In the following we present the generic method used in this work to obtain the residual contributions to the particle current. We calculate explicitly the mixed diffusive-sound contribution appearing in  Eqs.~\eqref{eq:resJx}, i.e.,  
\begin{widetext}
\be
\label{eq:RdTex}
\begin{split}
\int_{\bm K}\,
\sin\phi_{\bm K}\,
\mc R^{(r)}_{d{\bm T}_1}&=\,
\int_{\bm K}\,
\mc E^x\,\sin^2\phi_{\bm K}\,
\left\{
\left[\int_{\tau_0}^\tau\,d\tau'\,\frac{\mc G_{d{\bm T}_1}(\tau',\tau,{\bm K})}{\bw(\tau')}\,\left(\mc R_{{\bm T}_1{\bm T}_1}+\mc R_{dd}\right)\right]
\,-\,\frac{\chi_{d{\bm T}_1}}{\bw(\tau)}\frac{1}{(D+\gamma_\eta)\,{\bm K}^2} \right\}\,,\\
&=\frac{\chi_{d{\bm T}_1}}{8\pi^2\,\bw(\tau)}\,
\mc E^x\,
\left(\frac{\sqrt{\pi}}{4}\,K_{d{\bm T}}^3\,\tau\,\mc F(\tau_0/\tau)-\,2\,\frac{\Lambda}{(D+\gamma_\eta)} \right)\,,
\end{split}
\ee
\end{widetext}
with $\mc F(\tau,\tau_0)$ given by 
%
\be
\label{eq:Ffunc}
\mc F(\tau_0/\tau)=\,\int_{\tau_0/\tau\to 0}^1\,\frac{dt}{t^{1/3}}\,
\int_0^\pi\,d(\cos\theta_{\bm K})\,\frac{1}{\left[B(t,\theta_{\bm K})\right]^{3/2}}\,.
\ee
%
In Eq.~\eqref{eq:RdTex} we replace the asymptotic solutions for $\mc R_{{\bm T}_1{\bm T}_1}$ and $\mc R_{dd}$ which for our purposes is enough since one is interested in the long time asymptotic behaviour. The integrand in Eq.~\eqref{eq:Ffunc} is divergent since $B(t,\theta_{\bm K})\approx 1-t$ when $t\to 1$~\cite{Akamatsu:2016llw}. The finite and linear divergent pieces of the integral~\eqref{eq:Ffunc} are obtained by rewriting it as follows
\begin{widetext}
\be
\label{eq:Ffunc2}
\begin{split}
\text{Eq.~\eqref{eq:Ffunc}}
&=\,\int_{\tau_0/\tau}^1\,\frac{dt}{t^{1/3}}\,\int_0^\pi\,d(\cos\theta_{\bm K})\,\left[\frac{1}{\left[B(t,\theta_{\bm K})\right]^{3/2}}-\frac{1}{(1-t)^{3/2}}\right]\,,\\
&+\,\int_{\tau_0/\tau}^1\,\frac{dt}{t^{1/3}}\,\int_0^\pi\,d(\cos\theta_{\bm K})\,\frac{1}{\left[1-t\,+\left(\frac{K_{d{\bm T}}}{\delta}\right)^2\right]^{3/2}}\,,
\end{split}
\ee
\end{widetext}
The integral in the first line is fully convergent while the second one is not and thus, an arbitrary cutoff parameter $\delta$ was introduced in order to regulate the divergence. The former regularized integral gives an analytical expression which can be Taylor series expanded and its the leading order $\mc O(\delta/K_{d{\bm T}})$ contribution read as 
\begin{widetext}
\be
\label{eq:regint}
\begin{split}
\int_{\tau_0/\tau}^1\,\frac{dt}{t^{1/3}}\,\int_0^\pi\,d(\cos\theta_{\bm K})\,\frac{1}{\left[1-t\,+\left(\frac{K_{d{\bm T}}}{\delta}\right)^2\right]^{3/2}}&\approx\,4\,\frac{\delta}{K_{d{\bm T}}}\,-\,\frac{63}{880}\,\frac{\Gamma\left(14/3\right)}{\Gamma\left(13/6\right)}\\
&+\,\text{suppressed finite terms}\,,\\
&\approx\,\,4\,\frac{\delta}{K_{d{\bm T}}}\,-\,1.72474
\end{split}
\ee
\end{widetext}
with $\Gamma(x)$ being the Gamma function. The supressed finite terms are of $\mc O\left((K_{d{\bm T}}/\delta)^n\right)$ with $n\geq 1$ which are suppressed. Now the numerical value of the finite piece of the integral~\eqref{eq:Ffunc2} is 
\be
\label{eq:Ffuncfin}
\lim_{\tau_0/\tau\to 0}\,\mc F(\tau_0/\tau)=-1.90286
\ee
By using Eqs.~\eqref{eq:regint} and \eqref{eq:Ffuncfin} into Eq.~\eqref{eq:RdTex} we finally get
\begin{widetext}
\be
\label{eq:RdTex-2}
\begin{split}
\int_{\bm K}\,
\sin\phi_{\bm K}\,
\mc R^{(r)}_{d{\bm T}_1}&=\frac{\chi_{d{\bm T}_1}\,\tau}{8\pi^2\,\bw(\tau)}\,
\mc E^x\,
\left(-\frac{1.60744}{\left[\left(D+\gamma_\eta\right)\tau\right]^{3/2}} +\frac{\sqrt{\pi}}{(D+\gamma_\eta)\tau}\delta -2\frac{\Lambda}{(D+\gamma_\eta)\tau}\right)\,,
\end{split}
\ee 
\end{widetext}
Therefore we conclude that the linear divergence cancels exactly with the associate late time divergence in Eq.~\eqref{eq:RdTex} iff
\be 
\delta=\frac{2}{\sqrt{\pi}}\,\Lambda\,.
\ee
This simple relation resembles the usual $\overbar{MS}$ renormalization scheme~\cite{Collins:1984xc}. 

By using the same regularization procedure for each integral in Eq.~\eqref{eq:resJ} we calculate the residual hydrodynamical fluctuating contribution to the current $J^\mu$. Below we list the numerical values of the relevant integrals needed in this calculation 
\begin{widetext}
\bs
\label{eq:listint}
\beal
\int_{{\bm K}}\sin\theta_{\bm K} \,\cos\phi_{\bm K}\,\left(\mc R^{(r)}_{++}-\mc R^{(r)}_{--}\right)&=-4\frac{(1-v_a^2)}{v_a}\,\frac{\mc E^x\,\tau}{\bw}\,\frac{1}{(\gamma_\eta\tau)^{3/2}}\,(0.0133)\,,\\
-\int_{{\bm K}}\cos\theta_{\bm K}\,\cos\phi_{\bm K}\,\mc R^{(r)}_{d{\bm T}_2}&=-0.01522\,\frac{\tau\,\chi_{d{\bm T}}}{\bw}\,\frac{\mc E^x}{\left[\left(D+\gamma_\eta\right)\tau\right]^{3/2}}\,,\\
\int_{{\bm K}}\cos\theta_{\bm K}\left(\mc R^{(r)}_{++}-\mc R^{(r)}_{--}\right)&=-\,4\frac{(1-v_a^2)}{v_a}\,\frac{\mc E^\varsigma\,\tau}{\bw}\,\frac{1}{\left(\gamma_\eta\,\tau\right)^{3/2}}\,(0.02141)\,,\\
\int_{{\bm K}}\sin\theta_{\bm K}\,\mc R^{(r)}_{d{\bm T}_2}&=-0.008\,\frac{\tau\,\chi_{d{\bm T}}}{\bw} \frac{\mc E^\varsigma}{\left[\left(D+\gamma_\eta\right)\tau\right]^{3/2}}\,,
\end{align}
\es
\end{widetext}
The sum of the different contributions give rise to the numerical values of the particle current quoted in Eq.~\eqref{eq:finJ}. For the residual contribution of $\Delta J^\tau$~\eqref{eq:deltaJt} we used the results listed in Table 1 of Ref.~\cite{Akamatsu:2016llw}. 
\bibliography{hydrokinchempot}

\end{document}